\documentclass[pdflatex,sn-mathphys-num]{sn-jnl}
\AtBeginDocument{
\newgeometry{
  left=3cm,
  right=3cm,
  top=3cm,
  bottom=3cm
}
}

\usepackage[caption=false]{subfig}

\theoremstyle{plain}

\theoremstyle{definition}

\theoremstyle{remark}

\usepackage{xcolor}
\definecolor{pastelorange}{RGB}{255, 204, 153}
\definecolor{pastelyellow}{RGB}{255, 255, 153}

\usepackage{makecell} 
\usepackage{multirow}
\usepackage{multicol} 
\usepackage{booktabs} 
\usepackage{array} 
\usepackage[table]{xcolor} 
\usepackage{tabularx} 
\usepackage{caption}
\usepackage{comment}
\usepackage{algorithm}
\usepackage{algpseudocode}
\usepackage{amsmath} 
\usepackage{graphicx}
\usepackage{amssymb}  
\usepackage{amsfonts}

\begin{document}

\title{Influence of the majority group on individual judgments in online spontaneous conversations.}

\author*[1]{\fnm{Diletta} \sur{Goglia}}\email{diletta.goglia@it.uu.se}

\author[1]{\fnm{Davide} \sur{Vega}}\email{davide.vega@it.uu.se}

\author[1]{\fnm{Alessio} \sur{Gandelli}}\email{alessiogandelli99@gmail.com}

\affil[1]{\orgdiv{InfoLab, Department of Information Technology}, \orgname{Uppsala University}, \orgaddress{\street{Lägerhyddsvägen 1}, \city{Uppsala}, \postcode{SE-752 37}, \country{Sweden}}}

\abstract{
This study investigates how the majority group influences individual judgment formation and expression in anonymous, spontaneous online conversations. Drawing on theories of social conformity and anti-conformity, we analyze everyday dilemmas discussed on social media. First, using digital traces to operationalize judgments, we measure the conversations' disagreement and apply Bayesian regression to capture shifts of judgments formation before and after the group's exposure. Then we analyze changes in judgment expression with a linguistic analysis of the motivations associated with each judgment. Results show anti-conformity behaviors: individuals preserve the majority's positive or negative orientation of judgments but diverge from its stance, with persuasive language increasing post-disclosure. Our findings highlight how online environments reshape social influence compared to offline contexts.
}


\maketitle

\section{Introduction}\label{sec:intro}

Social norms delineate acceptable beliefs and behaviors within communities, fundamentally shaping social structures and human interactions~\citep{doi:10.1177/14614448241307035}. 
These norms guide both individuals' judgment \textit{formation}, influencing how they interpret and evaluate circumstances, and individuals' judgment \textit{expression}, determining when they feel comfortable or compelled to declare their judgments publicly~\citep{Bursztyn2020}. For example, in political discussions, people may refrain from expressing dissent when they perceive the position of a group as strongly opposed to their own~\citep{Guo2023}. 

In online environments, social norms are often blurry and ambiguous, and their interpretation is challenging due to limited contextual cues and anonymity~\citep{de_candia_social_2022}. Users often misperceive the predominant opinion, resulting in social phenomena like pluralistic ignorance and majority illusion~\citep{Scheper15052025}.
Pluralistic ignorance emerges when individuals privately reject a norm or opinion but assume (incorrectly) that others accept it, hence going along with it publicly~\citep{plur_ignorance}. Majority illusion refers to the individual perception of a particular norm or opinion to be more common than it actually is, often because it is held by highly visible users~\citep{Lerman_2016}. 
Despite these difficulties, individuals observe other users online and adjust their actions and thoughts accordingly, rationally and collectively constructing new shared rules and expectations. For this reason, digital platforms thrive the change of existing social norms and the creation of new ones. 

Within such a digital ecosystem, the majority of conversations are \textit{spontaneous}, i.e., they are naturally occurring and evolving, user-initiated, and not elicited by an external prompt. 
Spontaneous conversations emerge from users' intrinsic motivations to share and engage~\citep{gilmartinChatChunkTopicCasual2018, gilmartinJustTalkingModelling2018, blommaertConvivialityCollectivesSocial2015}, and evolve organically without external prompting or guidance~\citep{herring2004approach, prabhuDefiningQuantifyingConversation2020, Clift_Haugh_2021}, especially in self-organized online communities~\citep{kotutWindsChangeSeeking2022}. 
Spontaneous conversations encompass playful jokes, nonsenses, and discussions that span an open-ended spectrum of topics, from everyday life situations to complex social issues~\citep{gilmartinChatChunkTopicCasual2018}. For instance, a recent viral phenomenon spreading wildly across TikTok and Reddit is the ``Italian brainrot''~\citep{ujkItalianBrainrot2025}, consisting of AI‑generated pictures of creatures, generating a considerable amount of spontaneous discussions. 
Anonymity further reinforces the spontaneous nature of online conversations by lowering social and reputational constraints, with consequences in increasing users' self-disclosure~\citep{Clark-Gordon04052019,Anonymityinonlinecommunication}, for example in blog discussions~\citep{10.1111/jcc4.12008}.

Conversely, non-spontaneous conversations 
include orchestrated exchanges, which evolution is purposely and externally conducted and framed. Examples include online interviews, edited podcasts, Q\&A during press conferences, or discussions in Open Source Software (OSS) communities. While users in Reddit can freely decide whether to follow community guidelines (i.e., they maintain their spontaneity), OSS have highly guided and explicitly structured practices of governance~\citep{chakraborti2024we}, that highly affect both users' participation and conversational content, thereby undermining the spontaneity~\citep{HallinanEtAl2025Aspirational, Frderer2017SelfGovernanceOO, leskovec2010governance}.

The limited external regulation of spontaneous conversations reduces accountability, enabling participants to communicate more freely and often facilitating the formation of social ties among peers. Engagement in non goal-oriented exchanges, including discussions of trivial topics, fosters informal interaction. Consequently, the construction of social norms and the dynamics of social influence are also less constrained: examining spontaneous discussions can reveal patterns that would be less visible under more regulated conditions. By contrast, non-spontaneous interactions are typically goal-oriented and structured around objectives such as consensus-building or problem-solving. Their higher level of accountability usually requires participants to adhere to predefined social expectations, making such interactions less conducive to the formation of social ties and to unrestricted opinion expression. This distinction is important because the regulation of both the scope and the dynamics of interaction affects how social influence can emerge and be observed. For example, in an online interview, social influence is mediated by the controlled formulation, sequencing, and potential editing of questions and responses. These factors constrain and structure participation, making non-spontaneous conversations less suitable for the study of emergent norms.

Despite the increasing availability of spontaneous conversations, 
a considerable portion of the literature in computational social science studied human judgments formation and expression focusing on three main approaches. The first consists of the use of evidence collected from lab experiments and surveys~\citep{das2014modeling, CINNIRELLA20072011, 10.3389/fpsyg.2020.02254, Shin17062025, KIM2019109}. The second approach grounds the analysis direct interaction between single individuals~\citep{eagly1993psychology, griffin2006first, sherif1961social, sherif1965attitude}. The third approach narrows experiments to limited subject areas,
such as marketing and political debates (e.g., product recommendation, elections, propaganda speech, climate change;~\citep{novotnaIncivilityIntoleranceCOVID192023, 10.1145/3332168, messaoudi_opinion_2022, cortis_over_2021, Widmann_Simonsen_2025, Taylor27062024}), overlooking more informal topics of discussion.
A missing aspect in the literature is studying individual judgment expression in relation to an anonymous \textit{group} of users in \textit{spontaneous} discussions, a crucial social communication aspect to explain collective phenomena online~\citep{greve_2016}.

In this work we measure whether and to what extent individuals align their judgments with the majority group in online spontaneous conversations. We contribute to the understanding of how the exposure to an hidden online group influences individuals' judgment formation (what they think) and expression (how they say it).

Hence, our research question is the following:

\begin{quote}
\textbf{RQ:} Is the reveal of the majority (group) judgment affecting the expression of individual judgments in anonymous and spontaneous online discussions?
\end{quote}

We draw on established work in cognitive and social psychology concerning collective conformity (Section \ref{sec:theoretical_framework}) and we investigate the applicability of offline socially grounded theories to online interactions. We consider the two following competing hypotheses:

\begin{quote}
\textbf{H1:} When the group judgment is publicly disclosed, individual judgments collectively \textit{converge} towards it. Social norms of collective \textit{conformity} hold for online anonymous and spontaneous conversations.
\end{quote}
\begin{quote}
\textbf{H2:} When the group judgment is publicly disclosed, individual judgments collectively \textit{diverge} from it. The online scenario, favoring disinhibition, exerts a strong influence on individual behavior thus confirming the occurrence of \textit{anti-conformity} norms.
\end{quote}

We study the \textit{group} influence on individuals in anonymous and spontaneous conversations on social media platforms, focusing our attention on discussions around everyday dilemmas (Section \ref{subsec:data}). We start with assessing the individual exposure to the group by leveraging digital traces provided by the platform. These traces include explicit and unambiguous stances of the judgments expressed by users. We directly exploit such stances as the ground truth of both individual and majority judgment. In this way, we avoid introducing uncertainty, ambiguity, or approximation in quantifying user stances (Section \ref{subsec:measuring_judg_and_op}). 

Then, we measure the \textit{change} of both individual judgments' formation and expression after the exposure to the majority group.
To compare judgments' \textit{formation} before and after the group exposure, we compute the disagreement of a discussion as the multi-label entropy of the judgments expressed before and after. This measure represents a proxy for conformity effect (Section \ref{subsec:judgment_disagreement}).
To quantify the collective influence exerted on individual judgments, we employ a Bayesian multivariate regression approach (Section \ref{subsec:model}), conducting the analysis at the group level.  
Subsequently, we compare the distribution of judgments before and after the majority (Section \ref{subsec:rdd}) to further analyze the shifts in judgments expression after the exposure.
Finally, in order to examine judgments' \textit{expression}, we conduct a linguistic analysis on the motivations around individual judgments, expressed in the text (Section \ref{subsec:text-analysis}). We conclude by assessing the discrepancy in such expression following exposure to group influence.

Our results show that:
\begin{itemize}
    \item The exposure to the group judgment has a significant impact on individual judgments formation (Sections~\ref{subsec:ks},~\ref{subsec:assessment} and~\ref{rdd-results}). Users exhibit a systematic tendency to \textit{not} conform their judgment to the majority group (Sections~\ref{subsec:disagreement}). 
    Following the exposure, individual judgments preserve the group’s positive or negative orientation (Section~\ref{rdd-results}).

    \item The public disclosure of the majority judgment impacts individual judgments expression as well. 
    Regardless of what the majority is, after its disclosure, users continue to engage in discussion while expressing fewer judgments (Sections~\ref{subsec:assessment} and~\ref{rdd-results}). At the same time, we find a significant increase in linguistic patterns indicative of persuasive language, after the exposure to the majority (Section~\ref{subsec:social_dim}). Users whose judgments \textit{agree} with the majority are more likely to express opinions conveying trust, support, and knowledge. Users who \textit{disagree} with the majority are more likely to express opinions conveying similarity and power.
    
\end{itemize}

To conclude, we interpret our findings through the lens of the aforementioned theoretical framework (Section~\ref{conclusion}) and argue that online environments shape social influence mechanisms in ways that meaningfully differ from offline contexts.

\section{Theoretical framework}\label{sec:theoretical_framework}
In this section, we outline the theoretical framework used to guide the study and to interpret its results. We begin by introducing the main theories and concepts that ground this research and support our findings (Section \ref{sec:traditional}). Then, we differentiate between foundational work on conformity and anti-conformity, highlighting the key distinctions relevant to our analysis. Finally, we revisit our hypothesis in light of the theoretical perspectives discussed (Section \ref{sec:comp_studies}).

\subsection{Research on offline environments}\label{sec:traditional}

\paragraph*{Judgments vs. opinions.}
In sociology and social psychology, an \textit{opinion} is the expression of a belief~\citep{stephenson_perspectives_1965}, i.e., a general estimation of a target (e.g., a fact or a person) on a dimension ranging from negative to positive~\citep{sherif1965attitude, barbara_2023}. 
In contrast, a \textit{judgment} about something or someone requires people to engage in an inferential process: they evaluate and draw conclusions from some external evidence~\citep{apa_dict}. 
In cognitive psychology, the formation of a judgment is often studied as a Bayesian process in which prior beliefs are revised and updated in light of some new observable and verifiable information, to produce a posterior opinion~\citep{physics6030062, MACIEL2020124293}. 

Prior beliefs are relatively resistant to updates and enduring over time~\citep{Kahan_2013}, especially when related to the domain of morality and values~\citep{doi:10.1177/14614448241307035}. They are part of the self- (or ego-) system, derived from specific cultural contexts, emotions, and past behaviors associated with the target~\citep{olson1993attitudes}. As a consequence, if other people have the same information and values as we do, we expect them to agree with our judgments, otherwise 
conflicts, radicalization, and polarization~\citep{physics6030062, tajfel1979integrative} might result. 

These phenomena can lead to either conformity \textbf{H1} (people adjusting their views to align with others) or anti-conformity \textbf{H2} (people opposing and expressing diverging views).
The main goal of this work is to identify collective conformity or anti-conformity of users in relation to a majority group in online anonymous and spontaneous conversations.

\paragraph*{Individuals vs. groups.} Individual judgments formulated around moral values are closely linked to a \textit{group}~\citep{doi:10.1177/14614448241307035} (e.g., family, community, society) since they are always influenced by broader social forces such as culture, social class, and religion. 
Consequently, the influence of group dynamics on individual judgments' formation and expression has been a central focus in cognitive social psychology for decades. Foundational works include the Social Impact Theory (SIMT), the Social Judgment Theory (SJT), and the Social Identity Theory (SIDT). SIMT investigates how individuals are the source or the target of collective social influence, for example, through persuasive communication~\citep{latane1981psychology}. SJT studies how individuals evaluate new ideas comparing them with current attitudes~\citep{sherif1961social, chau_social_2014}. SIDT studies how individuals categorize themselves and others into groups, changing behavior towards both their own group and other groups~\citep{perdue1990us}.

Along with these theories, further notable literature include Moscovici's work and the Emergent Norm Theory (ENT). First, \cite{moscovici} demonstrated that minorities consistently expressing their viewpoint with confidence and coherence over time create doubt and internal conflict within the majority, eventually leading to private acceptance or even public change in opinion expression.
Second, \cite{turner1972collective} theorized ENT to understand the dynamic social process through which new norms are constructed in offline collectives. According to ENT, nontraditional behavior (i.e., all types of social behavior in which the conventional norms stop functioning as a guide) develops in groups as a result of the emergence of new conditions and circumstances~\citep{ent, Turner1996}. Specifically, the symbolic-interactionist perspective of ENT states that new norms
emerge through group spontaneous processes (without prior coordination) and develop through interactions (such as communication). As a consequence, anything that facilitates communication among groups' participants also facilitates the emergence of norms.

Recent studies in computational social science build upon all these foundational works by introducing a novel focus on digital environments. Nevertheless, much of the existing research in such field focuses on direct, individual-to-individual influence, rather than group-level influence.
Nowadays, social theories application to online platforms remains understudied: only a few works about communities of practice~\citep{Itao2025} build on ENT to understand the formation and evolution of social norms in digital spaces.
In this work we contribute to the understanding of how social norms are revised and how their expression shape the narrative of public discourse in online spaces, especially in spontaneous conversations.


\paragraph*{Conformity vs. anti-conformity.}\label{sec:confomity}
When people encounter differing opinions and judgments, their reactions can vary widely and polarize, ranging from conformity to anti-conformity. 

\textbf{Conformity} (also known as ``bandwagon'' effect) manifests when individuals tend to comply to the majority~\citep{nadeau1993new, Marsh_1985,fuBandwagonEffectParticipation2012, jadbabaie2022inference}.
This happens because of either normative or informational social influence. Normative influence occurs when individuals conform to avoid rejection and pursue approval and belonging~\citep{cialdini_conformity, spiral_of_silence}. For example, in Asch's experiment, participants conformed to the group’s incorrect answers to gain social acceptance. Informational influence takes place when individuals conform because the majority group behavior makes sense and they are rationally persuaded by the evidence~\citep{deutsch1955study}. 
A further example of conformism is ``internalization'', where the majority influences individuals because it is perceived as a credible and relevant source, with a behavior consistent with the individual’s value system~\citep{doi:10.1177/002200275800200106}.

\textbf{Anti-conformity} happens when the exposure to an opposing viewpoint strengthens individuals' pre-existing beliefs, leading them to adopt a minority view~\citep{maegherman2022law, choice_shift}.
Belief perseverance (also known as ``conceptual conservatism'') is the maintenance of a belief despite new information that firmly contradicts it~\citep{anderson2007belief}.
When beliefs are strengthened after an attempt to present evidence debunking them, we encounter the so-called ``backfire effect''. In social psychology, this refers to the unintended consequences of an attempt to persuade, resulting in the adoption of an opposing position. 

There have been a number of studies in the social sciences showing that the mere perception of belonging to two distinct groups is sufficient to trigger inter-group discrimination favoring the in-group. Even in anonymous and randomly-assigned groups, the presence of an out-group is sufficient to provoke competition, conflict, and polarization~\citep{tajfelIntegrativeTheoryIntergroup2001}.

\subsection{Research on digital environments}\label{sec:comp_studies}

Nowadays, most conversations happen online, involving invisible audiences and an increasing passive exposure to hidden individuals and groups. Social media indeed allows large populations to interact with random users from all over the world, fostering the tendency to behave differently online than in real life~\citep{mason2007situating, cheung2021online, Vilanova31122017} and leading to more unrestrained or uninhibited behavior. Two contributing factors are invisibility and dissociative anonymity. First, users feel less exposed, less accountable for their actions, and less concerned about consequences since their online identity is separate from their real-life identity~\citep{cheung2021online}. 
Second, where prior information about the participants is not available, users' accurate assessment of internal attitudes, group memberships, or prior beliefs becomes particularly challenging~\citep{IdentityandOnlineGroups}.

Both invisibility and dissociative anonymity contribute to the online disinhibition effect~\citep{suler_disinhibition, stuart2021measure, KIM2019109}, which promotes the development and spread of anti-social behaviors, such as cyberbullying, cyberharassment,
cyberaggression, and trolling~\citep{cheung2021online, Reicher_Spears_Postmes_Kende_2016}. 
Online disinhibition may significantly alter individual judgments' formation and expression, raising questions about whether theories developed in offline socially grounded contexts can be directly applied to online interactions. 


From the application of SIDT to computer-mediated communication (CMC), the social identity model of de-individuation (SIDE) originated~\citep{spears_lea_1994}, with the foundational idea that, during anonymous CMC, a user's social identity can be more or less \textit{salient}. In other words, an individual can identify more or less strongly with a reference group~\citep{terryGroupNormsAttitude1996}. When the social identity becomes salient, and the user identifies with a group, conformity to that group (due to normative influence) will be strong, even stronger than face-to-face interactions~\citep{trepteSocialIdentityTheory2006}. However, recent works have empirically contradicted this prediction~\citep{KIM2019109}.

Researchers continue to experience major computational challenges around accurately measuring online judgments, opinions, and their influence~\citep{Lerman_2016, battistella2019modelling, IdentityandOnlineGroups}, often leading to uncertainties, approximations, and oversimplified assumptions such as the binarization of stances~\citep{svetlana_2024, barbara_2023}. 
In most computational methods, opinions operationalization is too often too naive and not enough nuanced: 
research remains bound to qualitative methods to obtain the closest approximation to real opinions of users. In the present work we overcome such limitations by using digital traces extracted by the platform, hence preserving the ground truth of judgments and opinions expressed by users.  

Along with all these challenges, literature continues to overlook the pivotal role that \textit{spontaneous} conversations around values and norms plays on social media~\citep{doi:10.1177/14614448241307035}. Studies on the influence of the majority on individuals have predominantly addressed contexts involving issues of limited personal relevance to individuals, leading to superficial changes in opinions or behaviors~\citep{Capuano2024}. Only a minority of these works target norms and values, and those that do are conducted only on political studies~\citep{Aramovich01012012} or in lab settings~\citep{KIM2016241, Kundu01102013, https://doi.org/10.1111/jasp.12672}, overlooking spontaneous conversations. Research has only partially addressed whether majority influence extends to deeper individual value systems like social norms, and understanding whether traditional theories extend to these domains in the digital environment remains an open problem~\citep{Capuano2024}.

Studying conformity and anticonformity in spontaneous conversations is crucial, as these settings capture naturally emerging expressions of social acceptance and rejection under the absence of experimentally elicited interactions or external constraints. Organic informal online exchanges reflect how individuals actually and naturally position themselves relative to prevailing group norms, hence are more likely to reflect unsolicited influence, offering a ecologically valid understanding of such social processes.

In this work, we analyze spontaneous conversations from a Reddit community with the aim of verifying the occurrence and applicability of such foundational works to online environments.
In line with conformity theories, in the community, we could observe a convergence towards an agreement after the majority judgment is revealed to the participants (\textbf{H1}). This would suggest a shift towards group adherence despite the anonymous online settings weakening subjective norms (as supported by SIDE).
Conversely, according to anti-conformity theories, an increase in disagreement could happen in the community if users are deviating from the majority (\textbf{H2}) since, in anonymous online settings, the desire to be liked is less and people are increasingly uninhibited, losing their accountability.

\section{Data and methods}
\label{sec:methodology}

The primary focus of this work is to compare individual judgments expressed in spontaneous online conversations before and after the exposure to a majority (most popular) judgment. We achieve this by analyzing discussions within a Reddit community where users voluntarily share their thoughts and judgments on morally ambiguous everyday situations (Section~\ref{subsec:data}).
We download over 6,000 threads (i.e., post and related comments) and for each thread we measure both (individual and majority) judgments and (individual) opinions (Section~\ref{subsec:measuring_judg_and_op}). Specifically, we extract (i) individual judgments expressed by each user at the time of their comment, and (ii) the majority judgment disclosed by the platform. We measure opinions by detecting and quantifying the dimensions of 
communicative action in comments (Section \ref{subsec:text-analysis}). Next, we calculate the disagreement among users in each thread (Section~\ref{subsec:judgment_disagreement}) to analyze its evolution over time: this is motivated by the fact that a change in threads' disagreement could represent a proxy for the presence of conformity or anti-conformity effects~\citep{banish}.
We construct a Bayesian multivariate model (Section~\ref{subsec:model}) to measure the change of individual judgments expressed before and after the majority judgment is revealed. Finally we perform a local threshold analysis to compare the judgments expressions before and after (Section~\ref{subsec:rdd}).
    
\subsection{Data}
\label{subsec:data}

We ground our analysis on data obtained from Reddit, a social media platform where users participate in self-governing and self-organized communities, known as subreddits~\citep{jamnik2017use, medvedev2019anatomy}, 
typically built around a specific topic or theme, and serving as a valuable resource for research on social norms and user behavior~\citep{shatz_2017, hintz_2022, botzer_analysis_2023, gogliavega}. Subreddits have specific rules and guidelines that prescribe forbidden and not recommended behaviors. For each subreddit, rules and guidelines are displayed on the side of the webpage in order to facilitate their visibility. Users are supposed to write posts and comments that align with rules and guidelines. Each post, along with its comments, forms a thread~\citep{medvedev2019anatomy}. However, we have previously shown that around half of the users choose to participate without following the guidelines~\citep{gogliavega}, hence preserving the spontaneity of the discussion.

Subreddits are usually moderated by designated users who may both establish rules and guidelines and ensure that participants adhere to them. Moderators' activity is limited to deleting posts or comments and banning users, if needed to enforce order and control. This differentiates Reddit from highly governed online communities, such as OSS (Section~\ref{sec:intro}).

The \texttt{r/AmItheAsshole} (\texttt{AITA})\footnote{\url{https://www.reddit.com/r/AmItheAsshole}} subreddit represents an invaluable source of codified social norms~\citep{de_candia_social_2022}. In the \texttt{AITA} subreddit, users share personal experiences that have ambiguous moral outcomes, seeking a judgment on whether they had an unacceptable behavior in the narrated stories (in terms of the community, they were behaving as ``assholes''). Such stories are written in posts and typically include detailed descriptions and relevant background information about other people involved. 

In the \texttt{AITA} community, participants are encouraged by the community guidelines\footnote{\url{https://www.reddit.com/r/AmItheAsshole/about/rules}} to provide explicit judgments to express their stance about the characters' behavior in the story (either about all of them or only about the author of the post). To express a judgment, users can use a predefined list of acronyms made available by the community rules and summarized in Table~\ref{tab:flair_acronyms}. Users should include only one of the available acronyms as part of their comment, the one corresponding to the judgment they want to express. 

The \texttt{AITA} guidelines also suggest that, along with the acronym, users should include in the comment a brief motivation for their judgment, as it might be helpful for other readers. It is also important to note that, although the primary purpose of the comments is to express judgments, it remains the responsibility of each user to comply with the guidelines. Hence, alternatively, some users still participate by writing comments just to discuss and without judging~\citep{gogliavega}.

\renewcommand{\arraystretch}{1.5} 
\begin{table}[h!]
    \centering
    \footnotesize
    \begin{tabular}{c|c|c|c}
        \hline
        \textbf{Acronym} & \textbf{\makecell{Corresponding\\judgment}} & \textbf{Directed to} & \textbf{\makecell{Moral behavior\\in the story}}\\
        \hline
        \textbf{YTA} or YWBTA & ``You are the Asshole'' & \multirow{2}{*}{\makecell{The main character\\(i.e., author of the post)}} & \textcolor{red}{Negative}\\
        \cmidrule{1-2} \cmidrule{4-4}
        \textbf{NTA} or YWNBTA & ``You are not the Asshole'' & & \textcolor{blue}{Positive}\\
        \hline
        \textbf{ESH} & ``Everyone Sucks Here'' & \multirow{2}{*}{All characters involved} & \textcolor{red}{Negative}\\
        \cmidrule{1-2} \cmidrule{4-4}
        \textbf{NAH} & ``No A-holes Here'' & & \textcolor{blue}{Positive}\\
        \hline
    \end{tabular}
    \caption{Acronyms provided by the \texttt{AITA} community. Users can choose the acronym that corresponds to the judgment they want to express (about one or more characters of the story) and write it as part of their comment. Acronyms in bold are also used by the platform to broadcast the majority judgment (i.e., the final verdict).}
    \label{tab:flair_acronyms}
\end{table}

The \texttt{AITA} subreddit uses Reddit’s integrated voting system (i.e, the upvote button) to allow participants to rate the judgments they agree with. Users should upvote comments containing the acronym they think is correct. On the contrary, expressing disagreement with downvotes is not allowed: the downvote button should only be used to report off-topics, spam discussions, and harassing comments, in order to help moderators job of content-regulation. 

The \texttt{AITA} rules establish an 18-hours waiting period before assigning the final verdict. Users are supposed to write and upvote comments within this time frame if they want their judgments to be considered for the verdict. As users upvote different comments, a consensus emerges over time, with one judgment gaining the majority of agreements as the collective decision. The final verdict is automatically computed by the \texttt{AITA} algorithm by summing up all the upvotes that comments containing each acronym have received. After this time window has passed, this judgment is then accepted as the final verdict and it is made public by assigning a flair to the post, i.e., a tag with the respective acronym. In other words, the platform publicly reveals the final verdict, which states whether the main character (i.e., the author of the post) or other characters of the story had an unacceptable behavior. For example, if the most upvoted comments are those containing NTA, then ``Not the Asshole'' will appear as the thread's final verdict. In essence, the final verdict is the judgment most users agreed with. 

After the eighteen-hour threshold, users can continue participating independently of whether they have done so before. However, the majority judgment is calculated only once and will not be influenced by any judgment written afterwards.

Given this particular framework upon which the community is built, the \texttt{AITA} subreddit represents a precious source of spontaneous online conversations.
Such conversations are still dependent of platform governance: they are shaped by affordances such as the embedded voting system and community-specific rules about judgment expression. The relevant distinction for our study is that judgments are voluntarily (and anonymously) expressed in organically emerging and growing discussions around everyday-related topics. Spontaneity is central to our experimental setting since it allows us to examine shifts in judgment expression under ecologically rich conditions of social influence that are cannot be reproduced in elicited settings.

In the \texttt{AITA} community, the explicit request for a judgment is a requirement of the subreddit, allowing researchers to study how humans express moral judgments on other people online through socio-linguistic features. Indeed, the comments contained in \texttt{AITA} threads offer the ground truth of what people voted for and often why. This motivates why the \texttt{AITA} community has received much attention in recent literature~\citep{botzer_analysis_2023, giorgi_author_2023}, being also the most viewed Reddit community for four years in a row, from 2020 to 2023~\citep{2023_reddit_recap}. 

We collect 6,366 threads from the \texttt{AITA} community containing a total of 6,372,251 comments using the PRAW\footnote{Python Reddit API Wrapper (\url{https://praw.readthedocs.io/en/stable/})} library. The dataset is publicly available on Zenodo~\citep{goglia_2024_13620016}, and details about the data collection process are explained in~\cite{gogliavega}.

Our data exploration revealed that, on average, approximately half of the comments do not include any judgment, while a small percentage of them contain ambiguous judgments. This ambiguity occurs when a user either (i) writes a comment misspelling the judgment acronym, or (ii) uses more than one acronym in their comments, making the judgment invalid for the final computation. Although an attentive reader might infer the user's actual judgment from the text, none of these cases are taken into account by the system when it calculates the verdict. In our analysis, we handle these cases in the following way: (i) we mark comments without judgments with the \texttt{no\_judgment} label, (ii) we retrieve the corresponding correct acronym from misspelled judgments by using regular expressions, (iii) we label the comments containing multiple different acronyms as \texttt{unsure} judgments. This allows us to distinguish between users who had no intention to participate and those who did not participate because their comments have been invalidated.

The final verdict is publicly displayed close to the thread's title, hence being easily visible for users who join the discussion after eighteen hours. For this reason, we assume almost certain exposure to the majority judgment.
As a consequence, users who participate in a thread after the eighteen-hour threshold have been exposed to the majority judgment, and such exposure could bias individual judgments expressed afterward. 

Since we are interested in estimating the influence of the verdict, we remove threads that lasted less than eighteen hours, and threads that had less than 50 comments written after such a threshold to ensure the robustness of the results. The total number of threads obtained for the inference model is then 4,695, with approximately 4 million comments.

\subsection{Measuring judgments and opinions}\label{subsec:measuring_judg_and_op}

In this work, we exploit both acronyms included in comments and the final verdict to directly and unambiguously operationalize, respectively, individual and majority judgment.
The final verdict disclosure 
eliminates users' uncertainty surrounding the majority judgment, hence preventing the possibility of pluralistic ignorance or majority illusion effects. The exposure to the majority judgment prevents these misperceptions by establishing the ground truth of what the major group stance actually is. The public availability of the majority judgment provides a visible group-level signal to participants, against which we will compare subsequent individual judgments expression.

As described in Section \ref{subsec:data}, \texttt{AITA}'s community guidelines encourage users to contribute by providing both a judgment and a textual explanation. We have previously shown that approximately 50\% of users include a judgment in their comments~\citep{gogliavega}: hence, half of the contributions comprise either text alone (\texttt{no\_judgment}), while the other half includes a combination of judgments and text. 

To analyze opinions, we perform a pragmatic analysis of language~\citep{barbara_2023} 
using the model developed by \cite{montiLanguageOpinionChange2022} to detect the ten dimensions of communicative action from conversational texts. These dimensions are: knowledge, power, status, trust, support, similarity, identity, fun, romance, and conflict.
This model is particularly suited for our study for two reasons. First, since we are interested in measuring conformity or anti-conformity behaviors, we focus on the interactional aspects of conversations. Secondly, a pragmatic analysis of communicative actions applies because the \texttt{AITA} subreddit has the epistemic goal of determining the moral rightness or wrongness of actions, which requires analyzing not just what is said (judgments), but how it is said and the intentions behind it (opinions).
Table~\ref{tab:terminology} summarizes the distinction between opinions and judgments, as well as their operationalization in this work.

\begin{table}[h!]
    \centering
    \footnotesize
    \begin{tabular}{m{2cm}|m{4cm}|m{3.5cm}|m{3cm}}
        \hline
        \textbf{Term} & \textbf{Definition} & \textbf{Operationalization} & \textbf{Values}\\
        \hline
        \textbf{Opinion} & Expression of personal beliefs, attitudes, or thoughts about something or someone~\citep{barbara_2023}. & Extracting social dimensions of intent from comments' text. & Knowledge, power, status, trust, support, similarity, identity, fun, romance, conflict.\\
        \hline
        \textbf{Judgment} & ``Reasoned opinion''~\citep{Howe_Krosnick_2022}, revised after additional evidence or information. & Digital traces (acronyms) extracted from threads. & \multirow{2}{*}{\makecell[{{l}}]{\textcolor{red}{YTA} (or \textcolor{red}{YWBTA}), \\ \textcolor{red}{ESH}, \textcolor{blue}{NAH}, \\ \textcolor{blue}{NTA} (or \textcolor{blue}{YWNBTA}),\\ unsure, none.}}\\
        \cmidrule{1-3}
        \textbf{Individual judgment ($J$)} & Judgment expressed by participants about one (or more) character(s). & Acronym included in comments. & \\
        \hline
        \textbf{Majority judgment ($V$)} & Final verdict publicly revealed by the platform. & The most upvoted judgment (acronym). & A subset of $J$: \textcolor{red}{YTA}, \textcolor{red}{ESH}, \textcolor{blue}{NAH}, \textcolor{blue}{NTA}.\\
        \hline
    \end{tabular}
    \caption{Distinction between opinions and judgments and how we measure them. For each term, we define the corresponding variable used in the analysis, the definition obtained from the literature, how we extract and measure the variable, and the possible values it can take. Negative judgments are indicated in red, while positive judgments are indicated in blue.}
    \label{tab:terminology}
\end{table}
\renewcommand{\arraystretch}{1} 

\subsection{Computing disagreement}
\label{subsec:judgment_disagreement}

Disagreement, or the lack thereof, plays a crucial role in the formation of groups' opinions and judgments. Individuals expressing their judgments significantly influence collective decision-making processes, shaping the dynamic of the group's ability to reach a consensus~\citep{Oh2024Functional}. An increase or decrease in disagreement represents a proxy for detecting conformity or anti-conformity effects. Hence, as a first step, we aim to measure whether a change in the average disagreement in discussions occurs after the majority judgment has been publicly revealed. 

We expect to observe collective either conformity or anti-conformity behaviors from users. Judgments expressed afterward can reveal either a generalized agreement (bandwagon) or disagreement (backfire) towards the majority judgment.

In order to measure disagreement of \texttt{AITA} threads, we utilize judgments expressed in the comments as they represent the different stances that users are taking. We measure the level of disagreement of a thread by computing the proportion of all the stances taken by users in comments and by assessing the uncertainty of observing such stances in a thread. Inspired by \cite{de_candia_social_2022}, who aggregated different acronyms in a binary category (positive or negative) to measure the binary entropy on such aggregation, we opt for a multi-label entropy to operationalize disagreement (since we aim at including all the different stances expressed). 
We achieve this by computing the probability of each judgment appearing and measuring the Shannon entropy of a thread.
Given the set of acronyms (see Table~\ref{tab:flair_acronyms}) $J$, the entropy of a thread $T$ is defined as: 
\begin{equation}\label{eq:entropy}
    H_T(J) = - \sum_{j \in \mathcal{J}} p(j) \log p(j)
\end{equation}

where $p(j)$ is the discrete probability distribution of the judgments appearing in a thread's comments.
We rescale entropy values inn the range $[0, 1]$.
Values of the entropy close to 1 indicate maximum uncertainty and therefore maximum divisiveness: judgments are uniformly expressed, meaning participants equally take all the different stances. In this case, we can say that the thread has high disagreement. In contrast, a value of 0 represents the maximum level of certainty: all judgments are unanimous, and participants all agree on taking one stance, indicating that the thread has no disagreement.

We compute the entropy and update the total for every new comment added in the thread to analyze the evolution of disagreement among all threads (Figure~\ref{fig:disagreement}). Then, to obtain the variation over time and to obtain comparable variations at each timestamp, we round entropy values to the same unit of time (every minute).

Because disagreement is operationalized as a measure of stance distribution, we additionally compute the average polarization across all threads to compare the two measures and assess the robustness of the observed entropy patterns.
Let \(i\) be a thread and \(t\) a comment-time index within that thread, we define an indicator for whether comment \(t\) expresses one of the four stance labels. Then the polarization measure is the cumulative percentage of stance-bearing comments up to time \(t\):

\begin{equation}\label{eq:polarization}
    P_i(t) = 100 \times \frac{\sum_{\tau = 1}^{t} \mathbf{1}_{i\tau}}{t}
\end{equation}

\[\mathbf{1}_{it} =
\begin{cases}
1, & \text{if comment } t \text{ has acronym in } $J$ \\
0, & \text{otherwise}
\end{cases}
\]

\subsection{Bayesian inference model}
\label{subsec:model}

We model individual judgments based on the acronyms included in comments (see Table~\ref{tab:flair_acronyms}). First, we examine their distribution aggregated at user-level, modeling the judgment expression of each participant as a vector containing all the judgments they expressed for each post. We find that only 1\% of users participate again after the final verdict, expressing a new acronym in a new comment. This confirms that almost all the judgments expressed after the verdict are written by new users joining the discussion, rather than from users who already participated, confirming that users in Reddit often participate once~\citep{gogliavega}. For this reason, we assume individual judgments to be independent and identically distributed (i.i.d.).

We model the collective expression of such judgments in each thread as the distribution of each acronym appearing in the comments. For example, for a thread $T_i$, the judgment before and after could be represented by vectors $B_i = [.8, .1, .1, 0, 0, 0]$ and $A_i = [.5, 0, .1, 0, 0, .4]$. These vectors describe the percentage of, respectively, the judgments [``ESH'', ``NAH'', ``NTA'', ``YTA'', ``unsure'', ``no judgment''], and how much they changed after the verdict (see Appendix~\ref{app:judg_distribution}).
We aim to assess if and how much these distributions change in relation to the verdict disclosure (i.e., its influence on the judgment expression). To this end, we model our \textbf{RQ} as an inference problem through a multivariate linear regression approach. This allows us to simultaneously account for multiple variables 
and to assess their collective impact on the judgments after the verdict. We use a Gaussian linear model with weak informative prior distributions (Algorithm \ref{alg1}). We condition the predictor to be associated with the average change of the outcome (shift of individual judgments) after the verdict. 

Given the vector of possible judgments $J =$ [``ESH'', ``NAH'', ``NTA'', ``YTA'', ``unsure'', ``no judgment''], for each judgment $j$ in $J$, we run the following model,
\begin{equation}\label{eq:1}
    \mu_i = \alpha_{V[v]} + \beta (B_{j}-\bar{B})V[v] \quad \forall j \in J
\end{equation}

where:
\begin{itemize}
\item $\alpha$ 
represents the average judgments' deviation \textit{after} the verdict $i$ is acknowledged by users.
\item $V = [1, 2, 3, 4]$ is a vector encoding each possible verdict $v$ (``ESH'', ``NAH'', ``NTA'', and ``YTA'') as an integer. We intentionally do not consider the value 0 to avoid the case in which the prior will imply that $\mu$ for a verdict is more uncertain (before seeing the data) than $\mu$ for other verdicts.
\item $B_j$ represents the judgment $j$ expressed \textit{before} learning the verdict.
\item $\beta$ is the global model coefficient for variable $B_j$, representing the deviation from the mean $\bar{B}$ of the judgment $j$ after the verdict, due to the average change in the judgment before.
\end{itemize}

We obtain $|J|$ different models that estimate the impact of each verdict $V_v$ on each judgment $j$ expressed after the verdict disclosure.
We assume variables $B_j \quad \forall j \in J$ to be i.i.d.
We center the variable $B_j$ to reduce multicollinearity (correlation between predictors and their interactions) and to improve the numerical stability and interpretability of the models, with the result of improving their convergence and sampling efficiency, which is especially relevant when using MCMC methods.
For each model, we stratify by $V$ to allow the model to account for the influence of each single verdict $v$ separately. The inference is conducted within each stratum, estimating different parameters for each single verdict.

\paragraph*{Temporal robustness check.}
To assess whether the observed post-verdict disclosure patterns could also appear at comparative pre-disclosure points in time, we repeat the analysis using alternative ``fake'' thresholds. We shift the verdict threshold to periods in which no influence should be expected (since the verdict disclosure has not yet happened) and re-estimate the Bayesian models. This analysis, in combination with the KS test, provides a robustness check against unobserved and unexpected dynamics, such as underlying temporal trends or anticipation effects, further testing that the verdict influence is not an artifact of time-varying confounders.

\subsection{Local threshold analysis}\label{subsec:rdd}

The Bayesian model characterizes the influence of the verdict changes \textit{depending on the judgment expression before}, 
naturally due to the fact that post-verdict judgments are also conditioned on pre-verdict judgments (i.e., people might be influenced not only by the verdict reveal but also by what other judgments have been already expressed in the conversation).

In order to more closely approximate the verdict reveal influence on the participants, we focus the subsequent analysis on the time-related differences around the verdict threshold: given that the verdict reveal happens in a discrete point in time (always at 18 hours from the discussion start), we check for local discontinuities at the threshold time.

We implement a local linear comparison around the 18 hours temporal threshold to assess whether the judgment expression exhibits a discontinuous change at that time, thereby limiting pre-verdict judgments confounding. This approach provides a quasi-experimental test that is less dependent on global modeling assumptions, allowing us to better estimate the average post-verdict shift of judgments.

We run a local linear regression around the 18-hour threshold for every possible verdicts.
Then for each outcome \(Y_i\), the fitted model is the following:

\begin{equation}\label{eq:rdd}
Y_i = \alpha + \tau D_i + \beta_1 r_i + \beta_2 (D_i \cdot r_i) + \varepsilon_i
\end{equation}

which is equivalent to, 

\begin{equation}\label{eq:rdd_2}
Y_i =
\begin{cases}
\alpha + \beta_1 r_i + \varepsilon_i, & r_i < 0 \\
(\alpha + \tau) + (\beta_1 + \beta_2)r_i + \varepsilon_i, & r_i \ge 0
\end{cases}
\end{equation}

where, 

\begin{itemize}
    
    \item \(r_i\) is a running variable, equal to hours relative to the threshold \(t = 18\); 
    \(r_i=0\) is exactly the verdict time threshold. Only observations with \(|r_i| \le h\) are kept, with bandwidth \(h = 12\) hours.

    \item $i=1,2,3,...,n$ and $n$ is the number of discussion comments used in the regression window.
    
    \item \(Y_i\): the outcome at observation \(i\), $\forall j \in J$, where $J$ is a vector of percentage points.
    
    \item \(D_i = \mathbf{1}(r_i \ge 0)\): the post-threshold indicator. It is 1 on or after the threshold and 0 before it.
    
    \item \(\alpha\): the left-side intercept, i.e., the predicted outcome exactly at the threshold from the pre-threshold side.
    
    \item \(\tau\): the estimated discontinuity at the threshold, i.e., the intercept jump at \(r=0\). 
    
    \item \(\beta{_1}\): the pre-threshold slope.
    
    \item \(\beta{_2}\): the change in slope after the threshold.
    
    \item \(\beta{_1} + \beta{_2}\): the post-threshold slope.
    
    \item\(\varepsilon_i\): the error term.
    
\end{itemize}

This analysis holds attention to observations near 18 hours (within a 12-hours window before and after), to detect whether there exist a jump \(\tau\) in the outcome (judgment expression) at the threshold. The regression is estimated by Ordinary Least Squares (OLS) with HC1 robust standard errors, since an unweighted local linear fit is more suitable for our case study (the verdict is revealed at a discrete time stamp and does not have declining influence at the edges).

\subsection{Linguistic analysis of comments}\label{subsec:text-analysis}

We extract the topic of each thread from the post text by using BERTopic~\citep{grootendorst2022bertopic}. 
We leverage the topic analysis (Figure~\ref{fig:topic_distr} and Tables \ref{fig:topic_table} and \ref{tab:topic_examples}) to support the interpretation of our results (Section \ref{subsec:social_dim}) to ensure that extracted opinions and their evolution over time do not depend on specific discussion topics. 

Afterwards, we measure opinions expressed in comments' text by detecting and quantifying the ten dimensions of communicative action.
We run the Python implementation of the \texttt{tendimensions} model\footnote{\url{https://github.com/lajello/tendimensions}} for each of the 4M comments on a 4x NVIDIA Tesla V100 SXM2 GPU 32GB RAM server. 
The model consists of a multi-label classifier 
based on LSTM neural networks. It estimates the likelihood that a comment $c$ conveys a dimension $d$ by giving a score from 0 (least likely) to 1 (most likely). 
To facilitate the interpretation of the results, we binarize the returned scores to split 
comments between those that carry dimension $d$ with high probability and those that do not. Following the methodology of \cite{montiLanguageOpinionChange2022}, we do this via an indicator function that assigns dimension $d$ when it is above a certain threshold $\theta_d$. The use of dimension-specific thresholds is justified by the empirical distribution of the classifier scores varying across 
dimensions, making a fixed common threshold impractical. We take the value of $\theta_d$ as the 85th percentile of the empirical distribution of the scores.

We consider both text and acronyms only for comments expressing a valid judgment (i.e., we do not consider \texttt{no\_judgment} and \texttt{unsure} comments). To ensure a fair and robust comparison between opinions expressed before and after the verdict, we 
balance our dataset by selecting, for each thread, an equal number of comments before and after the verdict disclosure.
To assess the strength of the association between the opinion and the conformity (or anti-conformity) of judgments expressed in texts, we consider the odds ratios (OR) of finding dimension $d$ in comments agreeing with the majority verdict compared to those disagreeing with it. 
The OR between $d$ and the conformity of judgment only applies for comments written \textit{after} the final verdict disclosure, and are defined in Appendix~\ref{app:odd_ratios}.

\section{Results}\label{sec:results}

In this section, we present the results of our analysis. We begin with an evaluation of the consistency of judgments through a Kolmogorov–Smirnov test (Section~\ref{subsec:ks}), followed by an examination of the average thread disagreement and polarization over time (Section~\ref{subsec:disagreement}). Both these preliminary analyses were useful to run and evaluate the inference model (Section~\ref{subsec:assessment}). To conclude, we assess the change of judgment expressions (Sections~\ref{rdd-results} and~\ref{subsec:violin}) and opinions (Section~\ref{subsec:social_dim}) after the exposure to the majority.

\subsection{Comparing judgment behaviors before and after the verdict}\label{subsec:ks}

\begin{figure}[h!]
    \centering
    \includegraphics[width=.8\linewidth]{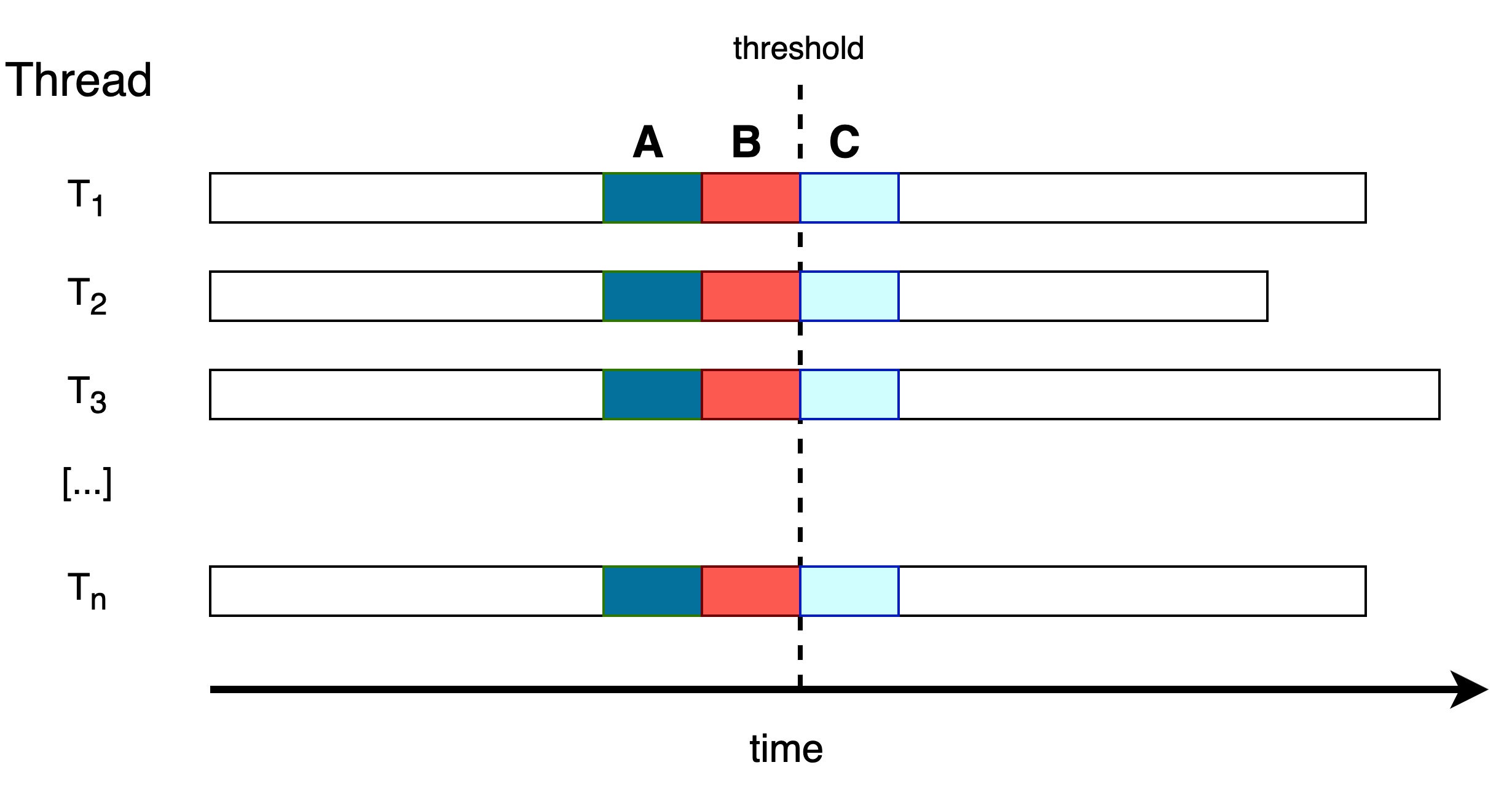}
    \caption{Experimental design of the judgments distribution comparison for the KS two-sample test. This has been performed for 800 threads ($n=800$). $A$, $B$, and $C$ intervals contains 50 comments each.}
    \label{fig:thread}
\end{figure}

To assess whether a a difference exists between judgments expressed before and after the majority, we compute the two corresponding distributions and compare them. This preliminary analysis allows us to both estimate the consistency of judgments near the time of the verdict disclosure, and to ensure the robustness of the computation presented in this work. We compare the distribution of judgments between two pairs of time intervals (having the same size) from a sample of 800 threads by executing a Kolmogorov–Smirnov (KS) two-sample test. Figure~\ref{fig:thread} illustrates the experimental design of such a comparison. For each thread $T_i$, we consider the vector representing the acronym distribution\footnote{As described in \ref{subsec:model} the vector could be, for example, $T_i = [30, 0, 0, 20, 0, 50]$ with each element indicating the percentage of, respectively, [``ESH, ``NAH'', ``NTA'', ``YTA'', ``unsure'', ``no judgment'']}. The first interval includes the distribution referring to the last 100 comments written before the verdict disclosure, and it is further split into two intervals of equal size ($A$ and $B$ in the figure). Then we create a third interval ($C$ in the figure) that includes the first 50 comments written after the verdict disclosure. We apply a KS test to compare the cumulative distributions of $A$ and $B$ intervals (both before the verdict), and then $B$ and $C$ (before and after the verdict, respectively), to determine whether there are significant differences in judgment distributions.

Table~\ref{tab:ks} shows the results of the KS test, which indicate a statistically significant difference between the judgments before ($B$) and after ($C$) the eighteen-hour threshold, suggesting that the majority judgment, disclosed after such threshold, has a notable impact on the subsequent judgments. 
This is further corroborated by the absence of significant differences between the two distributions before the eighteen-hour threshold ($A$ and $B$).

\begin{table}[h!]
    \centering
    \begin{tabular}{c|cc|cc}
          & \multicolumn{2}{c}{\textbf{After (B and C)}} & 
         \multicolumn{2}{c}{\textbf{Before (A and B)}}\\
         \hline
         Judgment & KS stat & p-value & KS stat & p-value \\
         \hline
         ESH & 0.27 & $\ll$ 0.0001 & 0.015 & 0.99 \\
         NAH & 0.27 & $\ll$ 0.0001 & 0.023 & 0.98 \\
         NTA & 0.27 & $\ll$ 0.0001 & 0.01 & 0.99 \\
         YTA & 0.14 & $\ll$ 0.0001 & 0.01 & 0.99 \\
         unsure & 0.31 & $\ll$ 0.0001 & 0.03 & 0.89 \\
         no judg & 0.38 & $\ll$ 0.0001 & 0.018 & 0.99 \\
    \end{tabular}
    \caption{Results of Kolmogorov–Smirnov two-sample test that compares the judgment distributions of $A$ and $B$ intervals (before the verdict), and $B$ and $C$ intervals (before and after the verdict).}
    \label{tab:ks}
\end{table}
\renewcommand{\arraystretch}{0.9} 

\subsection{Disagreement evolution over time} 
\label{subsec:disagreement}

\begin{figure}[h!]
    \centering

    \includegraphics[width=0.8\textwidth]{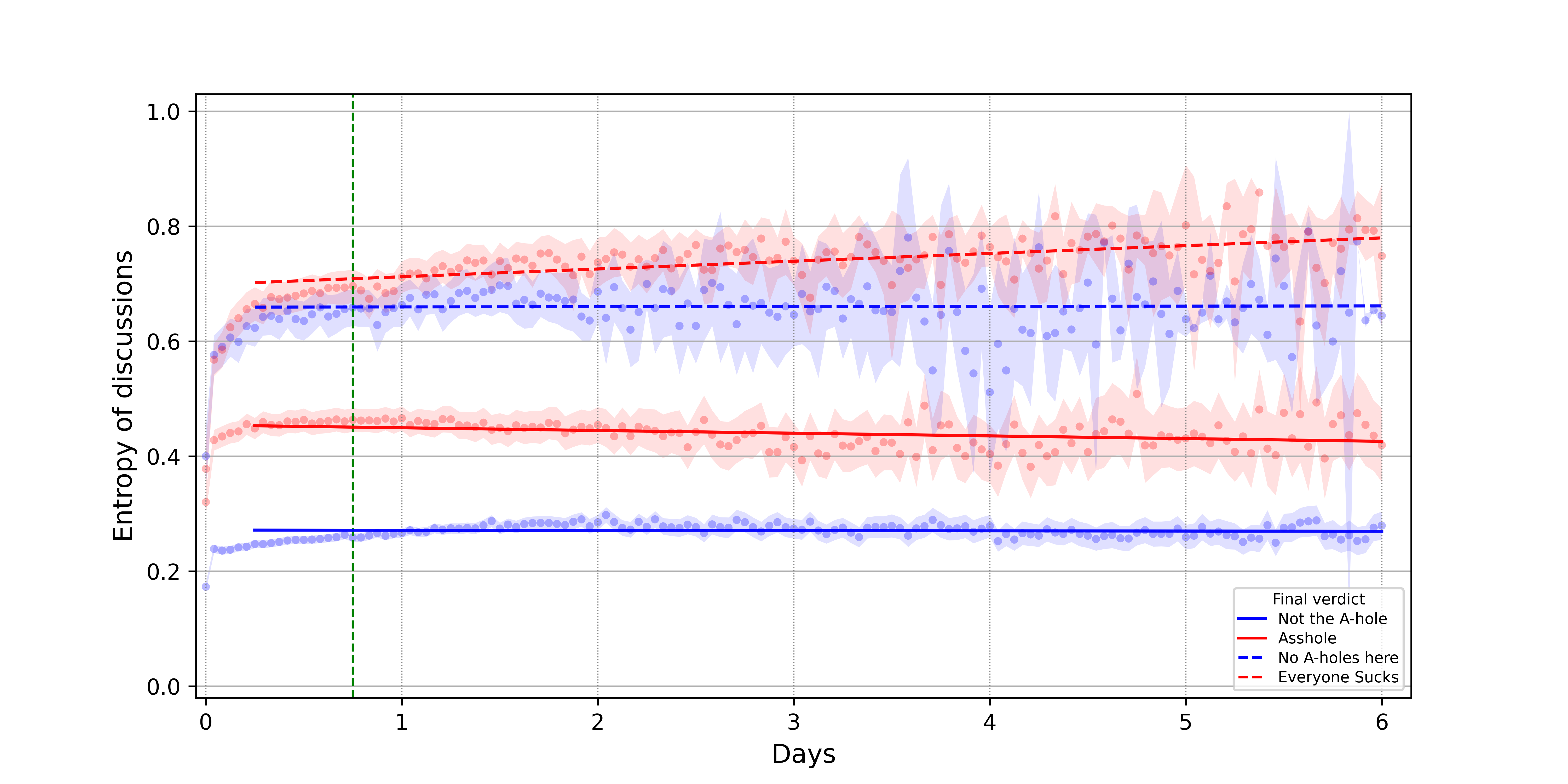}

    \caption{Disagreement (entropy) evolution over time, fitted over all discussions and grouped by final verdict. The dashed vertical line corresponds to the eighteenth hour, i.e., when the majority judgment is disclosed by the community and acknowledged by users.}\label{fig:disagreement}
    
\end{figure}

Figure~\ref{fig:disagreement} represents the evolution of disagreement over time, fitted over all threads and grouped by final verdict.
Negative verdicts (``ESH'' and ``YTA'') are represented in red, while positive verdicts (``NAH'' and ``NTA'') are represented in blue.
Solid lines correspond to verdicts related only to the author of the post (``NTA'' and ``YTA''), while dashed lines refer to verdicts that also involve other characters of the story (``NAH'' and ``ESH'').

Because entropy is inherently sensitive to thread size and temporal accumulation, early phases of a discussion can produce mechanically driven increases that do not reflect meaningful behavioral dynamics. In particular, when the initial distribution of stances is highly skewed, the subsequent appearance of additional categories can inflate entropy independently of any influence exerted by the majority opinion. Consistent with this concern, we observe that entropy trajectories typically exhibit an early ``elbow'' approximately after 5 hours discussion begins, marking a transition from rapid, composition-driven change to a more stable regime. To mitigate this source of bias, we restrict the entropy-over-time analysis to observations occurring after this point. In this way we the impact of initial variability and participation growth, ensuring that the measured entropy more reliably captures substantive evolution in stance distributions rather than mechanical effects. Furthermore, we compare this results with the polarization evolution over time (Figure~\ref{fig:polarization} in Appendix~\ref{app:polarization}) which corroborates the obtained disagreement patterns.

We can observe that all four curves corresponding to different verdicts do not significantly decrease after the majority judgment, hence suggesting the absence of a collective conformity effect towards it. Overall, threads' entropy after eighteen hours remains, on average, stable. In other words, learning the majority judgment has no substantial influence on reducing the disagreement of a discussion. Individual judgments do not collectively converge to an agreement with the group judgment.

The disagreement of individual judgments' formation is, overall, moderate or high. In order to analyze how such judgments are expressed, we compute the sentiment of each comment (Appendix~\ref{app:sentiment}), averaging it over all threads, and comparing the shift after the verdict. Sentiment has indeed been frequently employed as a proxy for disagreement~\citep{Hodel_2025, doi:10.1177/2056305120944393}, although it often provides an oversimplified representation of argumentative differences. In our result we find no relevant difference between the distribution of average thread sentiments before and after the verdict disclosure, confirming that disagreement is not necessarily expressed through a negative emotional tone: in healthy and constructive conversations, users articulate opposing views in a neutral or even positive way, which sentiment analysis alone may fail to capture accurately.

\subsection{Assessing the change of individual judgments after the majority judgment}
\label{subsec:assessment}

We examine the influence of the majority on individuals using the multivariate regression model described in Section~\ref{subsec:model}. 
Table~\ref{tab:model_result} summarizes the results of the analysis for all six models. The table also includes the 89\% interval boundaries of the posterior distribution and the diagnostics of the Markov Chain Monte Carlo (MCMC) model used for the inference.
Each row indicates a model parameter. $\alpha$ represents verdict-specific intercepts, $\beta$ represents verdict-specific slopes, while $\sigma$ indicates the standard deviation of models' residuals.
In the table's columns, \texttt{mean} is the posterior mean, \texttt{sd} is the posterior standard deviation, \texttt{hdi\_5.5\%} and \texttt{hdi\_94.5\%} are the 89\% Highest Density Interval (HDIs, also known as credible intervals), \texttt{mcse\_mean} and \texttt{mcse\_sd} are Monte Carlo Standard Errors of mean and standard deviation respectively, \texttt{ess\_bulk} and \texttt{ess\_tail} indicate the effective sample size (i.e., how many independent samples the posterior is equivalent to), \texttt{r\_hat} is the chain convergence diagnostic.
The model has a good performance and provides a reliable inference. The sampling noise is zero (indicating a precise estimate from the MCMC samples) and the sampling efficiency is substantial, indicating a good exploration of the posterior. \texttt{r\_hat} is 1 for all parameters, indicating an excellent model convergence across all chains.


Remarkably, all models show relatively small uncertainty 
and a meaningful impact of $\beta$, being always credibly different from zero. For significant $\beta$ parameters (bold in Table~\ref{tab:model_result}), the posterior uncertainty is relatively small in relation to the corresponding mean, indicating a low residual variability in the estimated values (i.e., narrow posteriors).
Note that, being the $\beta$ coefficient dependent on $V$, and being each verdict encoded as a positive scalar $V[v]$, the regression slopes referring to the same verdict are all rescaled by the same factor $V[v]$.

\paragraph*{Bayesian models' interpretation.}
The Bayesian regression estimates the individual judgment expression after the verdict as a function of both the verdict and the individual judgment expression before. Hence, the resulting measures in Table~\ref{tab:model_result} must be interpreted in terms of such global association.
The coefficient $\beta$ is linking the pre-verdict judgment expression to the post-verdict one after centering around the mean, measuring how much the outcome changes depending on the total influence of both variables $V$ and $B_j$.

As first result, we find that the baseline of ESH, NAH, and unsure judgments (i.e., intercepts $\alpha$ in Models 2, 3, and 5) suggests that they are unlikely to occur without other influencing factors. The corresponding distributions before the verdict (Figure~\ref{fig:judg_distr_after}) confirm a very low initial propensity from users to express these judgments.

Second, we observe that the disclosure of a \textit{negative verdict} has an influence on the expression of negative individual judgments NTA and NAH. 
The same holds for the opposite case: \textit{positive verdicts} influence the expression of positive individual judgments. 

Third, results show that expression of unsure judgments is not affected by any verdict (users being unsure about the judgment to express do not ``clear their mind'' after knowing what the majority is). The expression of no judgments is significantly affected by all majority judgments (i.e., by the verdict disclosure \textit{per se}, disregarding the type of verdict).

Finally, deviations of judgments after verdicts related to all characters ($\beta_{ESH}$ and $\beta_{NAH}$) show a wide range of values. In contrast, deviations after verdicts directed to the main character only ($\beta_{YTA}$ and $\beta_{NTA}$) have a narrower range. This observation aligns with the different level of disagreement of these two groups of verdicts (dashed versus solid lines in Figure~\ref{fig:disagreement}).

\newcommand{\boldunderline}[1]{\underline{\textbf{#1}}}
\newpage
\begin{table}
\captionsetup{font=footnotesize}
\centering
\footnotesize

\scalebox{0.9}{\begin{tabular}{lrrrrrrrrr}

\multicolumn{10}{l}{\textbf{Model 1) Judgment expressed: \textcolor{red}{YTA}}} \\
\midrule
\textbf{Param} & \textbf{mean} & \textbf{sd} & \textbf{hdi\_5.5\%} & \textbf{hdi\_94.5\%} & \textbf{mcse\_mean} & \textbf{mcse\_sd} & \textbf{ess\_bulk} & \textbf{ess\_tail} & \textbf{r\_hat} \\
\midrule
\textcolor{red}{$\alpha_{YTA}$} & 0.232 & 0.035 & 0.177 & 0.288 & 0.000 & 0.000 & 6083.0 & 5894.0 & 1.0 \\
\textcolor{red}{$\alpha_{ESH}$} & 0.340 & 0.078 & 0.216 & 0.465 & 0.001 & 0.001 & 6995.0 & 6802.0 & 1.0 \\
\textcolor{blue}{$\alpha_{NAH}$} & 0.183 & 0.100 & 0.018 & 0.341 & 0.001 & 0.001 & 6925.0 & 6434.0 & 1.0 \\
\textcolor{blue}{$\alpha_{NTA}$} & 0.138 & 0.102 & -0.028 & 0.297 & 0.001 & 0.001 & 6502.0 & 6128.0 & 1.0 \\
\textcolor{red}{$\beta_{YTA}$} &  \textbf{0.885} & 0.120 & \textbf{0.704} & \textbf{1.085} & 0.002 & 0.001 & 6344.0 & 6094.0 & 1.0 \\
\textcolor{red}{$\beta_{ESH}$} & \boldunderline{2.327} & 0.630 & \boldunderline{1.280} & \boldunderline{3.278} & 0.007 & 0.005 & 7684.0 & 6879.0 & 1.0 \\
\textcolor{blue}{$\beta_{NAH}$} & 0.478 & 0.688 & -0.587 & 1.602 & 0.008 & 0.006 & 6932.0 & 6359.0 & 1.0 \\
\textcolor{blue}{$\beta_{NTA}$} & 0.566 & 0.581 & -0.341 & 1.510 & 0.007 & 0.005 & 6477.0 & 6046.0 & 1.0 \\
$\sigma$ & 0.231 & 0.011 & 0.215 & 0.248 & 0.000 & 0.000 & 8788.0 & 5986.0 & 1.0 \\
\bottomrule\\[0.02ex]

\multicolumn{10}{l}{\textbf{Model 2) Judgment expressed: \textcolor{red}{ESH}}} \\
\midrule
\textbf{Param} & \textbf{mean} & \textbf{sd} & \textbf{hdi\_5.5\%} & \textbf{hdi\_94.5\%} & \textbf{mcse\_mean} & \textbf{mcse\_sd} & \textbf{ess\_bulk} & \textbf{ess\_tail} & \textbf{r\_hat} \\
\midrule
\textcolor{red}{$\alpha_{YTA}$} & 0.023 & 0.007 & 0.012 & 0.035 & 0.000 & 0.000 & 14543.0 & 7345.0 & 1.0 \\
\textcolor{red}{$\alpha_{ESH}$} & 0.025 & 0.029 & -0.022 & 0.072 & 0.000 & 0.000 & 11477.0 & 7917.0 & 1.0 \\
\textcolor{blue}{$\alpha_{NAH}$} & 0.006 & 0.019 & -0.024 & 0.038 & 0.000 & 0.000 & 14423.0 & 6928.0 & 1.0 \\
\textcolor{blue}{$\alpha_{NTA}$} & 0.005 & 0.007 & -0.007 & 0.016 & 0.000 & 0.000 & 13088.0 & 7719.0 & 1.0 \\
\textcolor{red}{$\beta_{YTA}$} & \boldunderline{0.365} & 0.166 & \boldunderline{0.089} & \boldunderline{0.621} & 0.001 & 0.001 & 14610.0 & 6199.0 & 1.0 \\
\textcolor{red}{$\beta_{ESH}$} & \textbf{0.429} & 0.237 & \textbf{0.042} & \textbf{0.794} & 0.002 & 0.002 & 11943.0 & 7529.0 & 1.0 \\
\textcolor{blue}{$\beta_{NAH}$} & 0.019 & 0.519 & -0.848 & 0.813 & 0.004 & 0.006 & 13902.0 & 6812.0 & 1.0 \\
\textcolor{blue}{$\beta_{NTA}$} & 0.135 & 0.216 & -0.218 & 0.472 & 0.002 & 0.002 & 13357.0 & 6459.0 & 1.0 \\
$\sigma$ & 0.075 & 0.003 & 0.070 & 0.081 & 0.000 & 0.000 & 13614.0 & 6675.0 & 1.0 \\
\bottomrule\\[0.02ex]

\multicolumn{10}{l}{\textbf{Model 3) Judgment expressed: \textcolor{blue}{NAH}}} \\
\midrule
\textbf{Param} & \textbf{mean} & \textbf{sd} & \textbf{hdi\_5.5\%} & \textbf{hdi\_94.5\%} & \textbf{mcse\_mean} & \textbf{mcse\_sd} & \textbf{ess\_bulk} & \textbf{ess\_tail} & \textbf{r\_hat} \\
\midrule
\textcolor{red}{$\alpha_{YTA}$} & 0.019 & 0.011 & 0.001 & 0.036 & 0.000 & 0.000 & 14054.0 & 7109.0 & 1.0 \\
\textcolor{red}{$\alpha_{ESH}$} & 0.015 & 0.044 & -0.059 & 0.083 & 0.000 & 0.000 & 8930.0 & 7493.0 & 1.0 \\
\textcolor{blue}{$\alpha_{NAH}$} & -0.009 & 0.039 & -0.067 & 0.056 & 0.000 & 0.000 & 11175.0 & 7192.0 & 1.0 \\
\textcolor{blue}{$\alpha_{NTA}$} & 0.043 & 0.012 & 0.023 & 0.061 & 0.000 & 0.000 & 12368.0 & 7677.0 & 1.0 \\
\textcolor{red}{$\beta_{YTA}$} & 0.417 & 0.212 & 0.075 & 0.744 & 0.002 & 0.001 & 15109.0 & 6521.0 & 1.0 \\
\textcolor{red}{$\beta_{ESH}$} & 0.349 & 2.258 & -3.201 & 4.043 & 0.024 & 0.021 & 9000.0 & 7399.0 & 1.0 \\
\textcolor{blue}{$\beta_{NAH}$} & \textbf{1.753} & 0.217 & \textbf{1.420} & \textbf{2.105} & 0.002 & 0.001 & 10614.0 & 7065.0 & 1.0 \\
\textcolor{blue}{$\beta_{NTA}$} & \boldunderline{2.065} & 0.420 & \boldunderline{1.405} & \boldunderline{2.740} & 0.004 & 0.003 & 11847.0 & 8121.0 & 1.0 \\
$\sigma$ & 0.115 & 0.005 & 0.107 & 0.123 & 0.000 & 0.000 & 12302.0 & 6886.0 & 1.0 \\
\bottomrule\\[0.02ex]

\multicolumn{10}{l}{\textbf{Model 4) Judgment expressed: \textcolor{blue}{NTA}}} \\
\midrule
\textbf{Param} & \textbf{mean} & \textbf{sd} & \textbf{hdi\_5.5\%} & \textbf{hdi\_94.5\%} & \textbf{mcse\_mean} & \textbf{mcse\_sd} & \textbf{ess\_bulk} & \textbf{ess\_tail} & \textbf{r\_hat} \\
\midrule
\textcolor{red}{$\alpha_{YTA}$} & 0.338 & 0.059 & 0.248 & 0.436 & 0.001 & 0.000 & 7681.0 & 6748.0 & 1.0 \\
\textcolor{red}{$\alpha_{ESH}$} & 0.356 & 0.098 & 0.199 & 0.512 & 0.001 & 0.001 & 6939.0 & 6266.0 & 1.0 \\
\textcolor{blue}{$\alpha_{NAH}$} & 0.555 & 0.074 & 0.432 & 0.670 & 0.001 & 0.001 & 9570.0 & 7585.0 & 1.0 \\
\textcolor{blue}{$\alpha_{NTA}$} & 0.402 & 0.036 & 0.348 & 0.462 & 0.000 & 0.000 & 7395.0 & 5748.0 & 1.0 \\
\textcolor{red}{$\beta_{YTA}$} & \boldunderline{0.984} & 0.235 & \boldunderline{0.606} & \boldunderline{1.356} & 0.003 & 0.002 & 7529.0 & 6404.0 & 1.0 \\
\textcolor{red}{$\beta_{ESH}$} & \boldunderline{1.225} & 0.658 & \boldunderline{0.200} & \boldunderline{2.299} & 0.007 & 0.005 & 7825.0 & 6574.0 & 1.0 \\
\textcolor{blue}{$\beta_{NAH}$} & \boldunderline{2.226} & 0.561 & \boldunderline{1.316} & \boldunderline{3.111} & 0.006 & 0.004 & 10036.0 & 7417.0 & 1.0 \\
\textcolor{blue}{$\beta_{NTA}$} & \textbf{0.811} & 0.133 & \textbf{0.611} & \textbf{1.033} & 0.002 & 0.001 & 7196.0 & 5499.0 & 1.0 \\
$\sigma$ & 0.240 & 0.011 & 0.221 & 0.256 & 0.000 & 0.000 & 12211.0 & 6879.0 & 1.0 \\
\bottomrule\\[0.02ex]

\multicolumn{10}{l}{\textbf{Model 5) Judgment expressed: unsure}} \\
\midrule
\textbf{Param} & \textbf{mean} & \textbf{sd} & \textbf{hdi\_5.5\%} & \textbf{hdi\_94.5\%} & \textbf{mcse\_mean} & \textbf{mcse\_sd} & \textbf{ess\_bulk} & \textbf{ess\_tail} & \textbf{r\_hat} \\
\midrule
\textcolor{red}{$\alpha_{YTA}$} & 0.004 & 0.002 & 0.001 & 0.008 & 0.000 & 0.000 & 11225.0 & 6554.0 & 1.0 \\
\textcolor{red}{$\alpha_{ESH}$} & 0.007 & 0.007 & -0.004 & 0.017 & 0.000 & 0.000 & 10745.0 & 7474.0 & 1.0 \\
\textcolor{blue}{$\alpha_{NAH}$} & 0.001 & 0.005 & -0.007 & 0.010 & 0.000 & 0.000 & 13548.0 & 6718.0 & 1.0 \\
\textcolor{blue}{$\alpha_{NTA}$} & 0.005 & 0.002 & 0.002 & 0.008 & 0.000 & 0.000 & 11175.0 & 7671.0 & 1.0 \\
\textcolor{red}{$\beta_{YTA}$} & 0.017 & 0.094 & -0.138 & 0.162 & 0.001 & 0.001 & 12012.0 & 7671.0 & 1.0 \\
\textcolor{red}{$\beta_{ESH}$} & -0.015 & 0.318 & -0.541 & 0.472 & 0.003 & 0.003 & 10432.0 & 7237.0 & 1.0 \\
\textcolor{blue}{$\beta_{NAH}$} & 0.030 & 0.596 & -0.896 & 1.004 & 0.005 & 0.006 & 11831.0 & 7817.0 & 1.0 \\
\textcolor{blue}{$\beta_{NTA}$} & -0.060 & 0.148 & -0.301 & 0.168 & 0.001 & 0.001 & 11282.0 & 7277.0 & 1.0 \\
$\sigma$ & 0.020 & 0.001 & 0.018 & 0.021 & 0.000 & 0.000 & 10565.0 & 7340.0 & 1.0 \\
\bottomrule\\[0.02ex]

\multicolumn{10}{l}{\textbf{Model 6) No judgment expressed}} \\
\midrule
\textbf{Param} & \textbf{mean} & \textbf{sd} & \textbf{hdi\_5.5\%} & \textbf{hdi\_94.5\%} & \textbf{mcse\_mean} & \textbf{mcse\_sd} & \textbf{ess\_bulk} & \textbf{ess\_tail} & \textbf{r\_hat} \\
\midrule
\textcolor{red}{$\alpha_{YTA}$} & 0.410 & 0.024 & 0.373 & 0.448 & 0.000 & 0.000 & 13478.0 & 7360.0 & 1.0 \\
\textcolor{red}{$\alpha_{ESH}$} & 0.334 & 0.083 & 0.201 & 0.466 & 0.001 & 0.001 & 9548.0 & 6984.0 & 1.0 \\
\textcolor{blue}{$\alpha_{NAH}$} & 0.213 & 0.064 & 0.107 & 0.310 & 0.001 & 0.000 & 12416.0 & 7477.0 & 1.0 \\
\textcolor{blue}{$\alpha_{NTA}$} & 0.376 & 0.021 & 0.345 & 0.412 & 0.000 & 0.000 & 12350.0 & 7298.0 & 1.0 \\
\textcolor{red}{$\beta_{YTA}$} & \boldunderline{0.934} & 0.150 & \boldunderline{0.696} & \boldunderline{1.172} & 0.001 & 0.001 & 13781.0 & 7405.0 & 1.0 \\
\textcolor{red}{$\beta_{ESH}$} & \boldunderline{2.438} & 0.804 & \boldunderline{1.131} & \boldunderline{3.701} & 0.008 & 0.006 & 9610.0 & 7299.0 & 1.0 \\
\textcolor{blue}{$\beta_{NAH}$} & \boldunderline{1.036} & 0.509 & \boldunderline{0.233} & \boldunderline{1.848} & 0.005 & 0.003 & 12592.0 & 7586.0 & 1.0 \\
\textcolor{blue}{$\beta_{NTA}$} & \boldunderline{0.915} & 0.157 & \boldunderline{0.653} & \boldunderline{1.155} & 0.001 & 0.001 & 12734.0 & 7077.0 & 1.0 \\
$\sigma$ & 0.241 & 0.011 & 0.224 & 0.258 & 0.000 & 0.000 & 10836.0 & 7400.0 & 1.0 \\

\hline\hline\\
\end{tabular}}
\caption{Summary of the posterior estimates for each model. $\alpha$ and $\beta$ coefficients correspond to the four possible verdicts. 
Significant changes of individual judgments after the verdict disclosure are \textbf{in bold} when they are the same as the majority, while \boldunderline{in bold and underlined} otherwise.}
\label{tab:model_result}

\end{table}

\begin{figure}[h!]
    \centering
    \includegraphics[width=\textwidth]{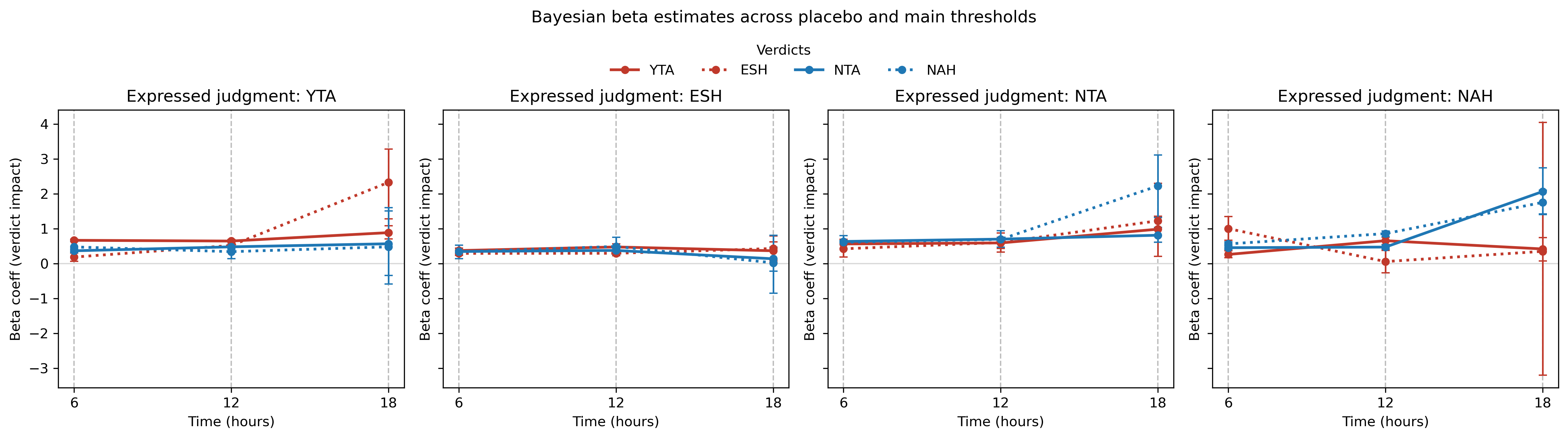}
    \caption{Results of the time placebo test. The $y$ axis shows the $\beta$ coefficient of the regression model, i.e., the shift from the average of the judgment expressed after the verdict.}
    \label{fig:placebo}
\end{figure}

\paragraph*{Temporal robustness check.}
We re-estimated the Bayesian regression models by artificially shifting the verdict to two timestamps prior to the actual verdict reveal. 
Figure~\ref{fig:placebo} shows that the estimated influence at placebo thresholds (i.e., 6 and 12 hours) are much smaller in magnitude than the real threshold (18 hours), supporting the validity of the significant influence of the verdict reveal on the subsequent judgment expression in the discussions. 
Overall, all the verdicts are associated with a stronger shift (increase or decrease) in the $\beta$ coefficient in correspondence to the real threshold, and not before it, thus confirming both the robustness of the previous analysis and the preliminary results obtained by the KS test (Section~\ref{subsec:ks}).

\subsection{Local threshold analysis}\label{rdd-results}
Regression discontinuities enables a more specific interpretation of the majority influence on individual judgment expression, since the verdict reveal is not entirely modulated by the prior judgment expression (see $\tau$ coefficient in Equation~\ref{eq:rdd}). The associated results, together with the evidence obtained by the Bayesian analysis, allow for a more complete assessment of how much the verdict disclosure is influencing subsequent judgment expressions.

Table~\ref{tab:rdd} in Appendix~\ref{app:rdd} reports the estimated discontinuity at the 18-hour cutoff for each verdict outcome, computed within a $\pm$12-hour bandwidth. At verdict time, we observe statistically significant discontinuities (despite small in magnitude) showing localized structural breaks, hence suggesting a discrete change of pre-existing judgments expressions after the verdict has been publicly revealed. 

\begin{figure}[h!]
\centering
\includegraphics[width=0.7\textwidth]{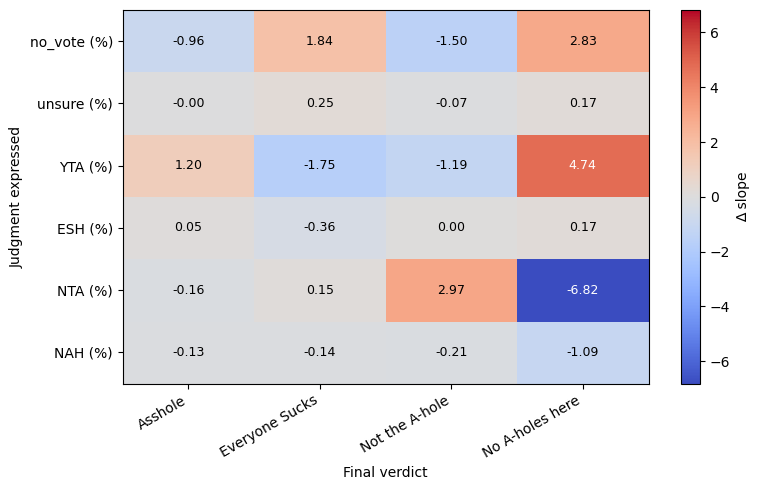}
\caption{Change of the regression slopes after the verdict reveal. Since raw slope coefficients in Table~\ref{tab:rdd} are expressed in percentage points per hour, we rescale them into the 12-hour movement to aid readability. Specifically, for each verdict-judgment couple, we compute the fitted percentage-point change associated with 12 hours of elapsed time on the corresponding side of the threshold. Positive values (red) indicate an increase of the judgment expression, whereas negative values (blu) indicate a decline. The symmetric color scale centered at zero allows direct comparison of the direction and magnitude of the trends across judgment-verdict groups.}\label{fig:rdd_heatmap}
\end{figure}

Figure~\ref{fig:rdd_all} in Appendix~\ref{app:rdd} shows the corresponding discontinuity patterns for each judgment across all four verdicts. The lines indicate the trend direction before and after the verdict disclosure.
Additionally, Figure~\ref{fig:rdd_heatmap} reports the difference between the regression slopes before and after the 18-hour threshold, separately for every judgment expressed and by final verdict. 
For the same judgment-verdict couple, red cells indicate a conformity effect, while blue cells indicate an anti-conformity effect. Conversely, for different judgment-verdict couples, red cells indicate anti-conformity. 
Within the 24-hour window symmetrically capturing both pre- and post-verdict disclosure, we can observe that:

\begin{itemize}
    \item Majority judgments directed towards the main character of the story (YTA and NTA) influence the post-verdict individual judgments in a conformity direction (see also Figure~\ref{fig:rdd_all} (a) and (d)). New participants express the majority judgment \textit{more} often than right before the verdict time, but continuing the same global trend (overall, the expression of such judgment decreases over time, in favor of more participants opting for ``no judgments''). 

    \item Conversely, majority judgments directed towards more than one character in the story (ESH and NTA) influence the post-verdict individual judgments in a non-conformity direction (see also Figure~\ref{fig:rdd_all} (b) and (c)). New participants express the majority judgment \textit{less} often than right before the verdict time, but again continuing the same global trend. NAH represent the utmost example since, after its public reveal, both a strong decrease towards positive judgments and a strong increase towards negative judgments are observed.

    \item Overall, the expression of no judgment always increases over time after the verdict reveal (Figure~\ref{fig:rdd_all} (f)), but at faster rate for verdicts directed towards more than one character in the story (ESH and NTA).

    \item In general, after the majority has been revealed, the distribution of judgments expression becomes much more scattered, suggesting that in a 12-hour window post-verdict the distribution of the four judgments (disagreement) has much more variability. There is not a global trend, across threads, of disagreement increasing or decreasing related to specific verdicts: the stance distribution changes differently for every discussion. As a result, we can observe an absence of \textit{collective} conformity or anti-conformity, coherently with what suggested, for a longer time period, by the analysis of disagreement (Section~\ref{subsec:disagreement}).
    
\end{itemize}


\subsection{Comparison of judgment distribution before and after exposure to the majority.}\label{subsec:violin}
\begin{figure}[h!]
    \centering
    \subfloat[Majority judgment: \textcolor{red}{ESH}]{\centering
        \includegraphics[width=0.9\textwidth]{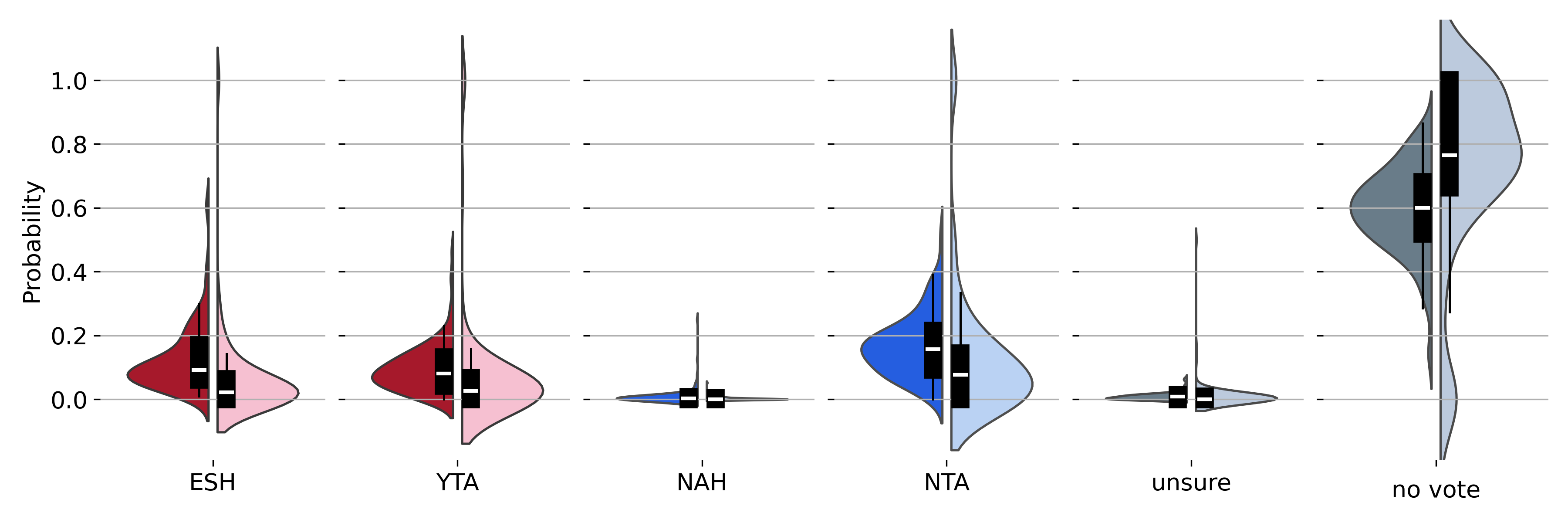}}

    \subfloat[Majority judgment: \textcolor{red}{YTA}]{
        \centering
        \includegraphics[width=0.9\textwidth]{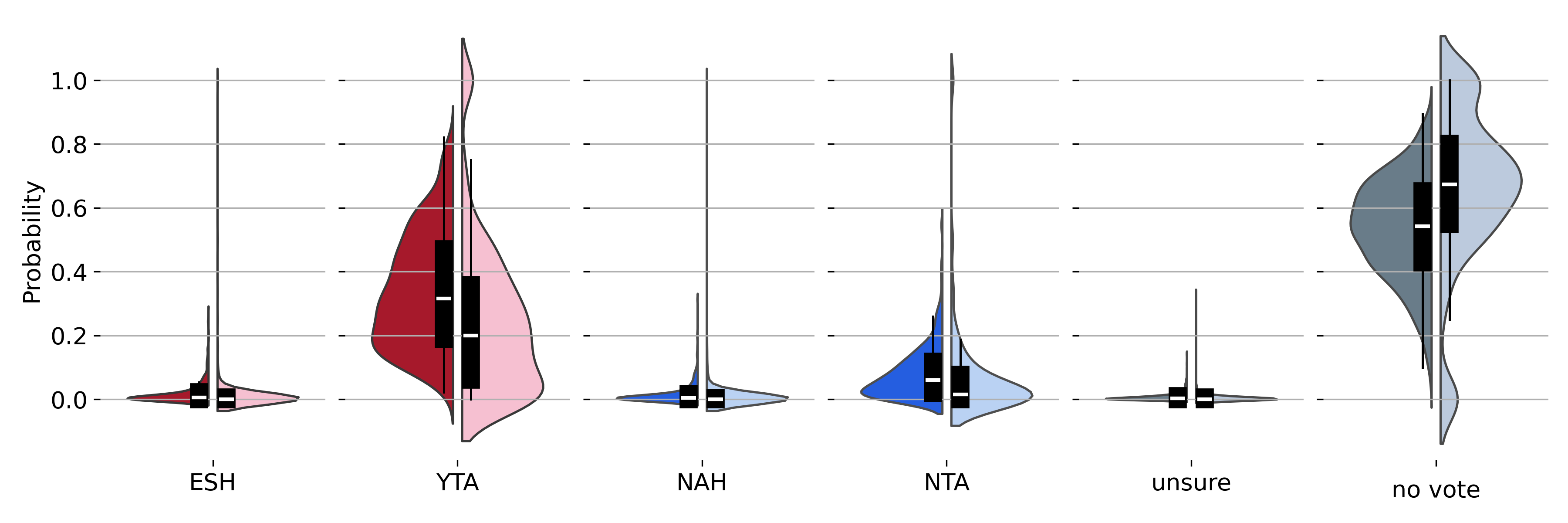}}

    \subfloat[Majority judgment: \textcolor{blue}{NAH}]{
        \centering
        \includegraphics[width=0.9\textwidth]{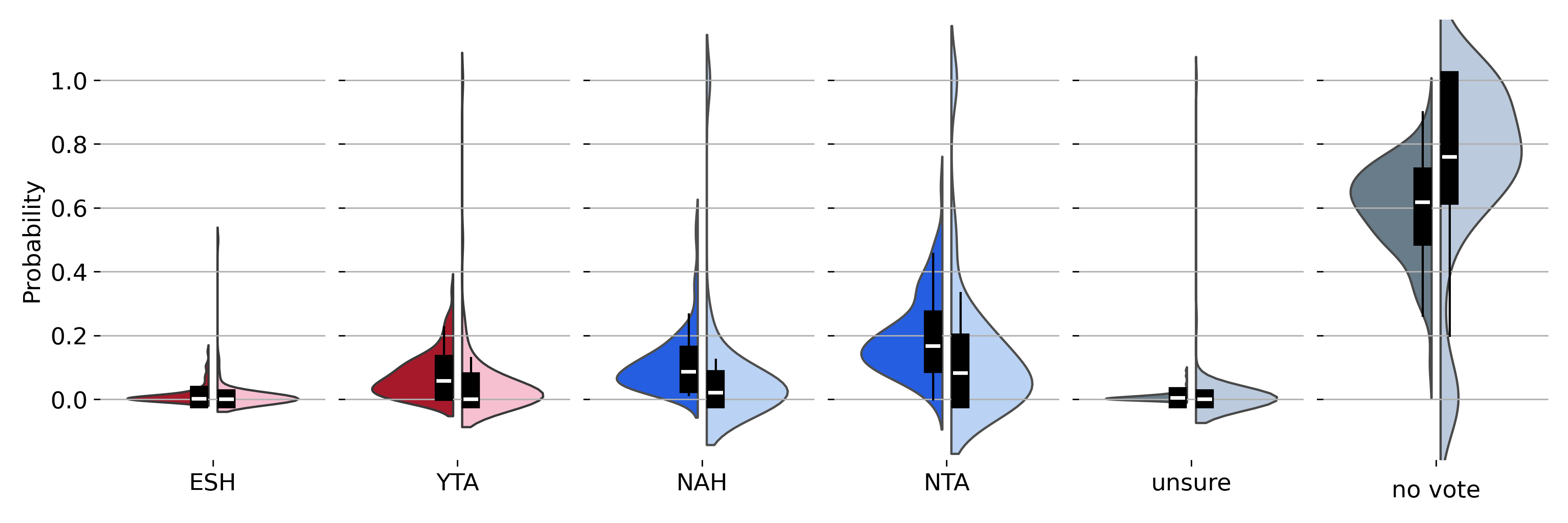}}
        
    \subfloat[Majority judgment: \textcolor{blue}{NTA}]{
        \centering
        \includegraphics[width=0.9\textwidth]{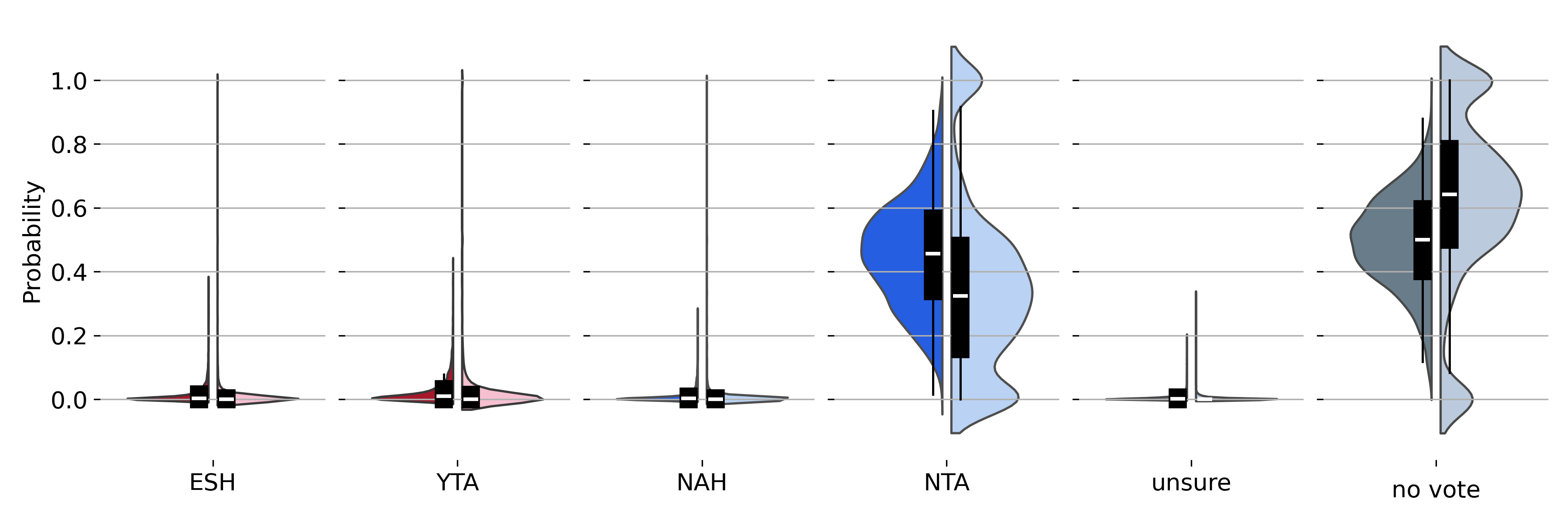}}
    \vspace{0.5cm}
    \caption{Comparison of the two distributions of individual judgments expressed in the comments before (left side, solid color) and after (right side, opaque color) the verdict acknowledgment.}
    \label{fig:violincomp}
\end{figure}

In order to further interpret models' results, we plot the comparison between the probability distributions of judgments before and after the verdict disclosure, grouped by each different verdict (Figure~\ref{fig:violincomp}).

First, we confirm that ESH, NAH, and unsure judgments have a remarkably low frequency both before and after the verdict disclosure.
Individual judgments involving all the characters of the story (ESH and NAH) did not constitute the largest part of judgments expressed before the majority judgment calculation (Figure~\ref{fig:violincomp} (a) and (c)). This confirms the high level of disagreement of the corresponding curves (dashed lines in Figure~\ref{fig:disagreement}) even before the eighteen-hour threshold.

Second, for all four different verdicts, judgment distributions after the verdict disclosure (right side in Figure~\ref{fig:violincomp}) increase their positive skewness (i.e., their tail extends to higher values). 
Globally, all means decrease in favor of the ``no judgment'' option, but new judgments are observed, occurring with a lower but non-negligible frequency since distributions reach extreme values. This increase of variability, coherently with the resulting $\beta$ parameters in Table~\ref{tab:model_result}, is the consequence of the verdict disclosure influencing individual judgments expressed afterwards. The verdict disclosure selectively impacts new users joining the discussions, driving them to express different judgments, hence amplifying the influence mechanism that pushes the values higher than they would have been before. This event globally reduces the magnitude of judgments expressed, but introduces new conditions (group influence on individuals) that push values to extreme levels (more diverse individual judgment expressions).

Consequently, the ``no judgment'' option has a wide distribution with a significant peak around mid-range probabilities, which moves to a higher range after the majority judgment disclosure. This indicates a relatively high frequency of non-expression of judgments (approximately 50\%, as illustrated in the preliminary analysis) that significantly increases after the verdict acknowledgment.

Overall, our findings suggest that the difference in the distributions before and after the verdict identified with the KS test (Section~\ref{subsec:assessment}) is mostly due to a systematic decrease of all the judgments expressed, in favor of comments containing no judgments.

\newpage

\subsection{Opinion expression after the verdict disclosure}
\label{subsec:social_dim}

\begin{figure}[h!]
    \centering
    \subfloat[Majority judgment: \textcolor{red}{YTA}\label{fig:sub_YTA}]{
        \includegraphics[width=0.47\textwidth]{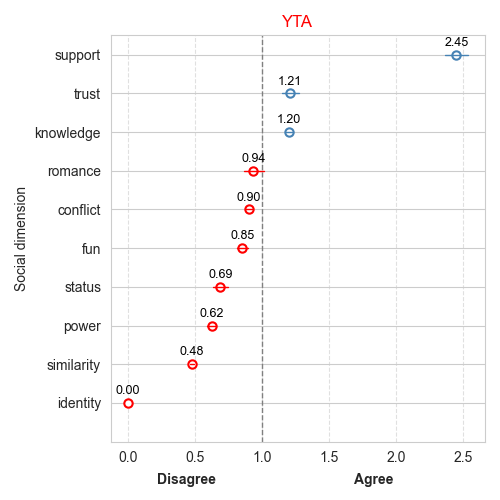}
    }
    \hfill
    \subfloat[Majority judgment: \textcolor{red}{ESH}\label{fig:sub_ESH}]{
        \includegraphics[width=0.47\textwidth]{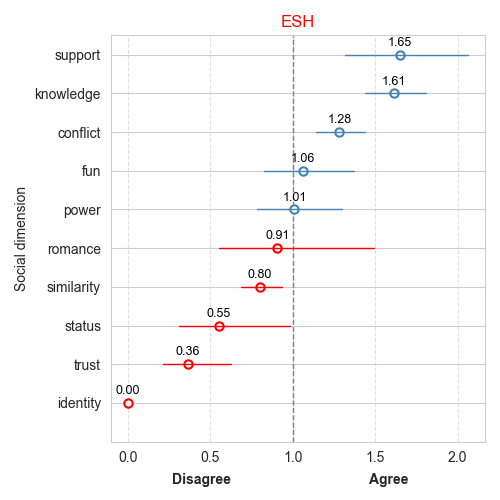}
    }

    \vspace{0.3cm}

    \subfloat[Majority judgment: \textcolor{blue}{NTA}\label{fig:sub_NTA}]{
        \includegraphics[width=0.47\textwidth]{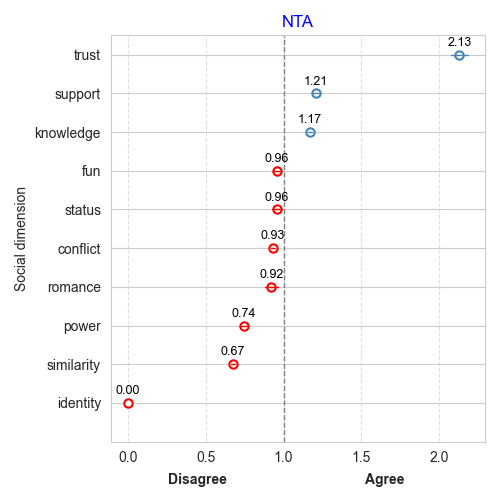}
    }
    \hfill
    \subfloat[Majority judgment: \textcolor{blue}{NAH}\label{fig:sub_NAH}]{
        \includegraphics[width=0.47\textwidth]{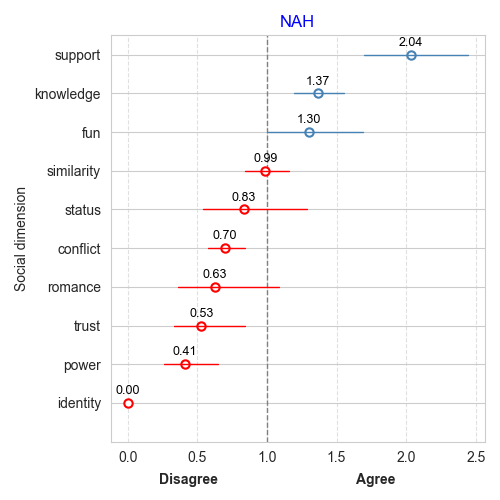}
    }

    \caption{Odd ratios of the ten social dimensions for each final verdict. Plots on the top (a and b) refer to negative verdicts, while plots on the bottom (c and d) refer to positive ones. Plots on the left side (a and c) refer to verdicts addressing only the main character of the story, while plots on the right side (b and d) address all the characters involved.}
    \label{fig:odd_ratios}
\end{figure}

Figures \ref{fig:odd_ratios} and \ref{fig:odd_ratios_for_each_verdict} illustrates the result of the textual analysis described in Section \ref{subsec:text-analysis}.
We find that all the ten dimensions have a significant association with the majority judgment being publicly revealed, with the exception of identity (i.e., the shared sense of belonging to the same group). 

The lack of this dimension in the dataset is not surprising. According to SIDE (Section \ref{sec:theoretical_framework}), one of the main differences between online and face-to-face communication lies in the way traditional markers of social identity (such as gender and age) are available or expressed~\citep{IdentityandOnlineGroups}. Especially related to individual identification in online communities, it has been shown how anonymity leads to de-individuation on the group level. In the \texttt{AITA} subreddit, conversations revolve around resolving interpersonal conflicts and seeking support, and participants are asked by the community rules to motivate their judgments by expressing their opinions on characters' behavior. Users are likely to answer the ``Am I The Asshole'' question appealing on private experiences, moral values, and personal wisdom, rather than constructing or asserting a group belonging with other participants or with the author of the story. This supports the obtained evidence about the absence of group identity expression in texts, already suggesting that we should not expect a confirmation of \textbf{H1}.

Overall, our results show that social intents are more likely to appear in comments that disagree with the majority, with the exception of trust, support, and knowledge (Figure~\ref{fig:odd_ratios}).
Users whose judgments \textbf{agree} with the majority are more likely to express opinions conveying trust (91\%) and support (46\%). This result confirms that trust is ``something we reserve mostly for those we already agree with''~\citep{MACIEL2020124293}. Trust and knowledge are two of the most important dimensions used for convincing arguments, i.e., to persuade someone~\citep{montiLanguageOpinionChange2022}. 
However, in the \texttt{AITA} subreddit, trust is used in agreement when the majority judgment refers to the main character only. When the majority judgment is directed towards all the characters of the story, trust is instead used in comments disagreeing with such majority (64\% more likely when the verdict is ESH and 47\% when the verdict is NAH).
Expressions of support and knowledge are the most used for all judgments that agree with the majority, both when this latter is positive or negative towards one or more characters.

Users whose judgments \textbf{disagree} with the majority are 37\% more likely to express similarity and 27\% more likely to express power. 
Expressions of similarity (i.e., communicating shared interests or motivations) and appeals to power are contributing to persuasive language as well~\citep{montiLanguageOpinionChange2022}. However, in the \texttt{AITA} subreddit, they are used in comments that disagree with the majority judgment.
This may be interpreted drawing on \cite{moscovici} foundational work (Section \ref{sec:theoretical_framework}) which states that minorities, in order to affect the opinion of a majority, should create a sense of connection with the majority: minorities who share similar interests or motivations with the majority may be more influential.
Along with this, another possible interpretation of this finding is that dissenters (i.e., users who identify with the minority since their judgments are disagreeing with the majority) might use similarity and power expressions to affirm and strengthen their ingroup status. However, since the measurement of saliency of participants social identity is out of the scope of this work, we cannot empirically assess this explanation and thus treat this interpretation as a plausible but unverified account to be examined in future research (Section \ref{conclusion}). 

Ultimately, our findings reveal no significant increase of the conflict dimension when users disagree with the majority group. This result indicates that conflict is not necessarily employed in language to express disagreement, contrary to expectations established in existing literature. This observation aligns with the null result obtained through sentiment analysis, suggesting that disagreement is not necessarily conveyed through negative sentiment or explicitly associated with conflict-related language.

We test the robustness of the opinion analysis by (i) computing the OR differentiated by topic, and (ii) conducting a qualitative assessment of the language model’s outputs to verify that identified dimensions are correctly attributed.
First, our results show that opinion expressions both in agreement and disagreement with the majority judgment do not change depending on the topic of the discussion. This consolidates the robustness of our analyses, proving that observed dimensions in opinion expression, whether aligning with or diverging from the majority, are a fundamental aspect of social interaction within the discussion, transcending the specific subject matter.  

Finally, we conduct a qualitative analysis of the opinions, by extracting the most representative 50 comments for each of the ten dimensions (i.e., containing the highest score). 
This subset was independently labeled by the three authors to establish whether a comment was directed to the author of the post or to some other users discussing in the thread. Based on this analysis, we conclude that all coders agree that the majority of comments between 80\% and 87\% of comments in the sample are directed toward the author of the post.
Hence, the change of dimensions refers to opinions about the author of the post.

\clearpage
\section{Discussion and conclusions}\label{conclusion}

In this work, we have analyzed anonymous and spontaneous online conversations in light of a revised social normative framework. Specifically, we examined whether the public disclosure of the majority judgment affects the expression of individual judgments. 
To answer our \textbf{RQ}, we found that in anonymous and spontaneous online discussions collected from the \texttt{AITA} subreddit, \textit{the public reveal of the majority judgment significantly affects the expression of individual judgments} (Sections~\ref{subsec:ks},~\ref{subsec:assessment}, and~\ref{rdd-results}).

In general, independently of what the majority judgment is, after its public disclosure, \textit{minority groups of users always emerge, expressing different judgments}. New users joining the discussion \textit{do not} collectively conform to the majority. Overall, anti-conformity trends across different discussions are stronger towards majority judging all the characters of the story (Sections~\ref{subsec:disagreement},~\ref{subsec:assessment}, and~\ref{rdd-results}).

After the public reveal of majority judgment, we also observed a shift the way in which judgments are expressed in the text (Section~\ref{subsec:social_dim}).

In summary, in the analyzed community:
\begin{itemize}
    \item Individual judgments are substantially affected by the acknowledgment of the majority judgment, but not collectively towards a conformity direction. 
    \item Anti-conformity individual judgments are more likely to be observed when the majority judges all the characters in the narrated story.
    \item Judgments expressed after the majority often preserve the positive/negative trend of the majority. 
    \item Users expressing judgments that do not conform with the majority motivate them in the text by appealing to similarity and power. In contrast, the minority of users agreeing with and conforming to the majority judgment, express support, knowledge, and trust in their comments. 


    \item Overall, publicly revealing the majority judgment to the community decreases the individual interest to explicitly judge their peers. 
    After acknowledging the verdict, new users joining a thread have a higher incentive to write comments and keep discussing the original post without expressing any judgment. 

    \item Regardless of the specific verdict, the majority has no global influence on whether a user remains in an ``unsure'' state regarding their judgment. This outcome can be attributed to the inherent ambiguity associated with these particular judgments, reflecting both users' initial hesitancy in forming a definitive stance, and the fact that the verdict was not useful to resolve their uncertainty. As further confirmation, we attested that instances of users commenting again to express a new judgment after the verdict disclosure are negligibly rare.
\end{itemize}

\paragraph*{Discussion.}
The observation of non-conformity (\textbf{H2}), despite it not being a collective trend, offers a novel and significant observational perspective on the dynamics of the emergence of norms in anonymous online spontaneous conversations. The disclosure of the majority judgment appears to not encourage convergence, probably driven by the unique affordances of digital communication, such as disinhibition and reduced social accountability, which collectively challenge the applicability of foundational theories. 

When interacting with their peers on online social media (especially if they are invisible or anonymous), users perceive a minor need for approval and
belonging than in the real world, lowering barriers to expressing non-normative views. As a result, users feel freer to articulate judgments that openly diverge from the majority, facilitating, on a broader scale, the transformation of values and the subsequent change of existing social norms and emergence of new ones~\citep{Turner1996}.

Our results contribute to observations of the emergence of distinct normative structures in digital spaces, highlighting the need to reconceptualize existing theories of social influence and norm change and formation when applied to digital environments.
The findings of this work invite interpretation through the lens of social theories on group influence. However, since direct measurements of relevant individual-level variables (e.g., the degree of identification with a group and the specific group with which individuals identify) falls outside the scope of this work, we cannot empirically evaluate these theoretical mechanisms within our current framework.

Future research could address this point by incorporating measures of social identity and group identification, enabling a more direct test of these theoretical explanations. For example, it would be valuable to investigate whether expressions of similarity in the \texttt{AITA} subreddit function as proxies for signals of group belonging, thereby partially compensating for the absence of the identity dimension. Moreover, future work could examine whether expressions of similarity and power are purposely employed to minimize intragroup differences and maximize intergroup differences.

\paragraph*{Limitations.}
Our analysis should not be interpreted as estimating the \textit{causal} effect of the majority verdict on individual judgments. Although the 18-hour disclosure threshold provides a temporal reference around which we observed significant shifts, post-verdict comments may differ from pre-verdict comments for several reasons, including user self-selection in content exposure and the platform’s algorithmic ordering of comment visibility in ways that are not directly observable. We therefore interpret our results as post-disclosure associations rather than as evidence that the verdict itself caused individual users to express specific judgments afterwards.

The findings of this study are most directly applicable to spontaneous conversations occurring in anonymous online settings. These environments meaningfully differ from other non-spontaneous online discussions where social influence and judgment formation occur, such as political votes, and where the expression of such judgments can have direct, real-world consequences for individuals. For instance, some public political statements may significantly impact people’s careers or freedoms. 
Consequently, the results of the present study may not be directly generalizable 
to non-spontaneous contexts with substantially different social dynamics.

Online spontaneous conversations, such as those examined in this research, often take on a playful or game-like character. Despite having potentially relevant interest for participants (for example, in shaping their self-presentation or gaining recognition within the group), comments in these forums rarely produce tangible consequences for others. In addition, depending on the online platform analyzed, specific topics under discussion may lead to different outcomes, since moral issues often elicit deeper value-based disagreements than the relatively lighthearted conversations examined here, which may limit the direct transferability of our findings.

Our results describe anti-conformity observed within the \texttt{AITA} conversational context, and may not directly extend to other online platforms where moderation, regulations, and guidelines differ.
These considerations are valuable for the computational social science research endeavor of rethinking ENT and its applicability to online settings, as digital environments foster social interaction that are qualitatively distinct from more consequential real-world contexts.

Ultimately, our Bayesian model represents the best available proxy for addressing our \textbf{RQ} with the available data. Nevertheless, it does not account for additional confounders that foundational works have identified as potentially relevant for social norms formation and change, such as individuals’ inner beliefs and prior knowledge~\citep{Turner1996}. While these dimensions fall outside the scope of our current analysis, acknowledging their importance is essential for situating our contribution within the broader theoretical landscape and for guiding future extensions of this line of research.

\bmhead{Supplementary information}
This manuscript contains supplementary information.

\section*{Declarations}

\subsection*{Availability of data and materials}
The data that support the findings of this study are openly available at Zenodo (\url{https://doi.org/10.5281/zenodo.13620016}) and on GitHub (\url{https://github.com/dilettagoglia/reddit-majority-opinion}).

\subsection*{Competing interests}
The authors have no competing interests to declare that are relevant to the content of this article.

\subsection*{Funding}

This work has been partly funded by eSSENCE, an e-Science collaboration funded as a strategic research area of Sweden. The computations were enabled by resources provided by the National Academic Infrastructure for Supercomputing in Sweden (NAISS), partially funded by the Swedish Research Council through grant agreement no. 2022-06725. The funders had no role in study design, data collection and analysis, decision to publish, or preparation of the manuscript.

\subsection*{Authors' contributions}

All authors contributed to the implementation of the research. 
D.G. contributed to data acquisition, design and developments of the methodology, software creation, validation, visualization, and formal analyses.
A.G. performed the topic modeling analysis.
D.G. and A.G. contributed to the original ideas of this work, as well as to the data and software curation. 
D.G. and D.V. contributed to the design of the methodology, as well as to the writing of the manuscript.
All authors read and approved the final manuscript.

\subsection*{Acknowledgements}
We thank Inga Wohlert for her insights and constructive feedback.

\newpage
\bibliography{bibliography}

\newpage
\appendix
\section{Supplementary material}

\vspace{1cm}

\subsection{Topic analysis}\label{app:topic_analysis}

The topic-related results presented in this study should be interpreted as exploratory and descriptive, providing an additional qualitative perspective on discussion structure. Unsupervised topic extraction based on embedding representations and clustering is indeed inherently sensitive to methodological choices (e.g., selection of the embedding model, clustering algorithm, parameter settings, preprocessing steps). As a consequence, the resulting topics do not correspond to a unique ground truth and may vary across alternative specifications. 
\vspace{0.5cm}

\begin{minipage}{\textwidth}

\centering
\begin{tabular}{r l r r r r}
\toprule
\textbf{Topic ID} & \textbf{Topic Label} & \textcolor{red}{\textbf{YTA}} & \textcolor{red}{\textbf{ESH}} & \textcolor{blue}{\textbf{NAH}} & \textcolor{blue}{\textbf{NTA}} \\
\midrule
0  & Family dynamics (female members) & 177 & 50 & 45 & 1330 \\
1  & Party                            & 84  & 16 & 27 & 570  \\
2  & Family dynamics (male members)   & 41  & 12 & 17 & 449  \\
3  & Eating                           & 50  & 12 & 9  & 291  \\
4  & Neighbor conflicts               & 22  & 9  & 4  & 177  \\
5  & Naming                           & 3   & 2  & 1  & 59   \\
-1 & Miscellaneous                    & 376 & 75 & 84 & 2313  \\
\bottomrule
\end{tabular}
\captionof{table}{Topics identified by the BERTopic model, with the corresponding final verdict size.}\label{fig:topic_table}
\end{minipage}

\begin{minipage}{\textwidth}
\centering
\includegraphics[width=0.9\textwidth]{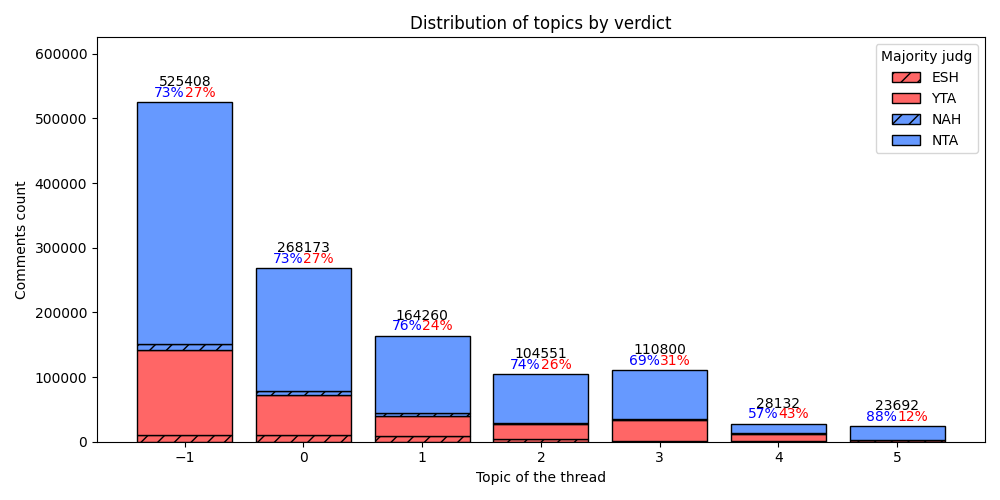}
\captionof{figure}{Distribution of threads' topics by final verdict (majority judgment)}
\label{fig:topic_distr}
\end{minipage}

\newcommand{\smallbullet}{\raisebox{0.1ex}{\scriptsize\textbullet}\hspace{0.5em}}
\begin{table}[h!]
\centering
\footnotesize
\begin{tabular}{p{0.5cm}p{2.2cm}p{10cm}}
\toprule
\textbf{ID} & \textbf{Topic Label} & \textbf{Examples (AITA for...)} \\
\midrule

\multirow{5}{*}{0} & \multirow{5}{*}{\makecell{Family dynamics\\(female members)}} & \smallbullet saying not again and not being happy for my daughter's pregnancy \\
 &  & \smallbullet laughing at my Ex and her husband for asking to have our daughter for another month \\
 &  & \smallbullet making a white woman cry \\
 &  & \smallbullet telling my friend her baby is the reason no one wants her around? \\
 &  & \smallbullet making my husband either take our son's ``shitbox'' or the bus because I will not lend him my car. \\

\midrule

\multirow{5}{*}{1} & \multirow{5}{*}{Party} & \smallbullet saying No To Dressing As A Disney Princess For A Wedding? \\
 &  & \smallbullet not showing up to my sister's wedding and calling her ungrateful? \\
 &  & \smallbullet failing to realize I wore a white blouse to a wedding? \\
 &  & \smallbullet telling my sister she will be insecure no matter what I wear to her wedding? \\
 &  & \smallbullet allowing only my twins at my wedding, but no other children? \\

\midrule

\multirow{5}{*}{2} & \multirow{5}{*}{\makecell{Family dynamics\\(male members)}} & \smallbullet telling my husband he needs to draw clear lines with the mother for his child? \\
 &  & \smallbullet telling my brother that I told him so and that his personality is the problem? \\
 &  & \smallbullet not apologizing to my brother for saying ``if he doesn't change his views he will die alone''? \\
 &  & \smallbullet threatening my brother to mortgage the house? \\
 &  & \smallbullet telling my father it's not my fault he failed at his dream? \\

\midrule

\multirow{5}{*}{3} & \multirow{5}{*}{Eating} & \smallbullet taking potatoes off a guy's plate at a wedding? \\
 &  & \smallbullet taking the largest slice of pizza because I paid for it? \\
 &  & \smallbullet asking someone why they expected gluten free options at a bread bakery? \\
 &  & \smallbullet refusing to go to a family event because I’d be pressured to eat food that goes against my dietary restrictions? \\
 &  & \smallbullet being rude about my veganism? \\

\midrule

\multirow{5}{*}{4} & \multirow{5}{*}{Neighbor conflicts} & \smallbullet telling all my parents' guests that my room has cannabis candy everywhere but they are still welcome to let their kids play in it. \\
 &  & \smallbullet asking people to be out of the gazebo that I paid to reserve at the park? \\
 &  & \smallbullet complaining and making the neighbor change their roof color? \\
 &  & \smallbullet refusing to give my stuffed animal to a baby? \\
 &  & \smallbullet calling the non-emergency line on my neighbor’s kid? \\

\midrule

\multirow{5}{*}{5} & \multirow{5}{*}{Naming} & \smallbullet ending a family naming tradition by not giving my son my late nephew's name? \\
 &  & \smallbullet ignoring people who called me by my ``old name'' \\
 &  & \smallbullet dumping my last name before a family member expires? \\
 &  & \smallbullet not gushing over the names she's picked out for her future twins... \\
 &  & \smallbullet not wanting my dad’s girlfriend's son to be referred to as my ``little brother''? \\

\midrule

\multirow{5}{*}{-1} & \multirow{5}{*}{Miscellaneous} & \smallbullet asking my husband to not eat lunch at night? \\
 &  & \smallbullet refusing to stop kissing my own baby? \\
 &  & \smallbullet not taking my youngest children on their weekend because my oldest daughter had a baby? \\
 &  & \smallbullet telling my son it’s absurd that he thinks we will be at his wedding \\
 &  & \smallbullet bluntly telling someone why their disabled son isn't allowed in my muscle car? \\

\bottomrule
\end{tabular}
\caption{Representative examples for each topic. For each topic ID, we extracted the five most representative threads' titles.}\label{tab:topic_examples}
\end{table}

\clearpage
\subsection{Judgments distributions}\label{app:judg_distribution}
\begin{figure}[h!]
\centering
    \subfloat[Before acknowledging the majority judgment.]{\centering
    \includegraphics[width=0.7\textwidth]{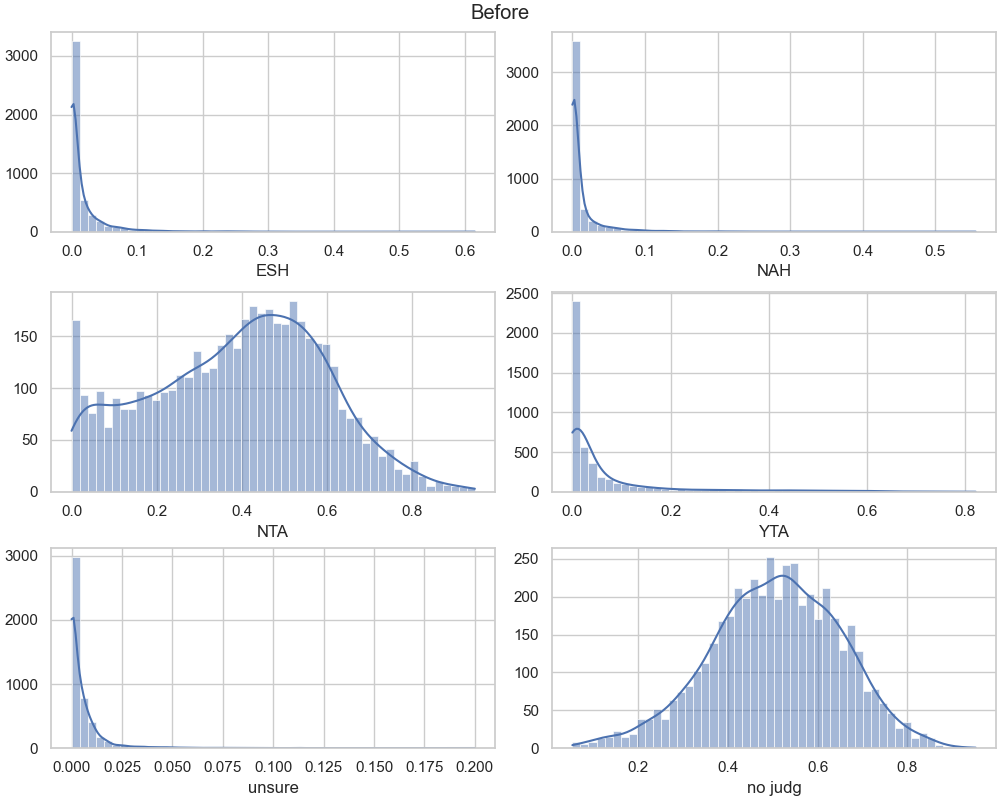}}

    \subfloat[After acknowledging the majority judgment.]{
    \includegraphics[width=0.7\textwidth]{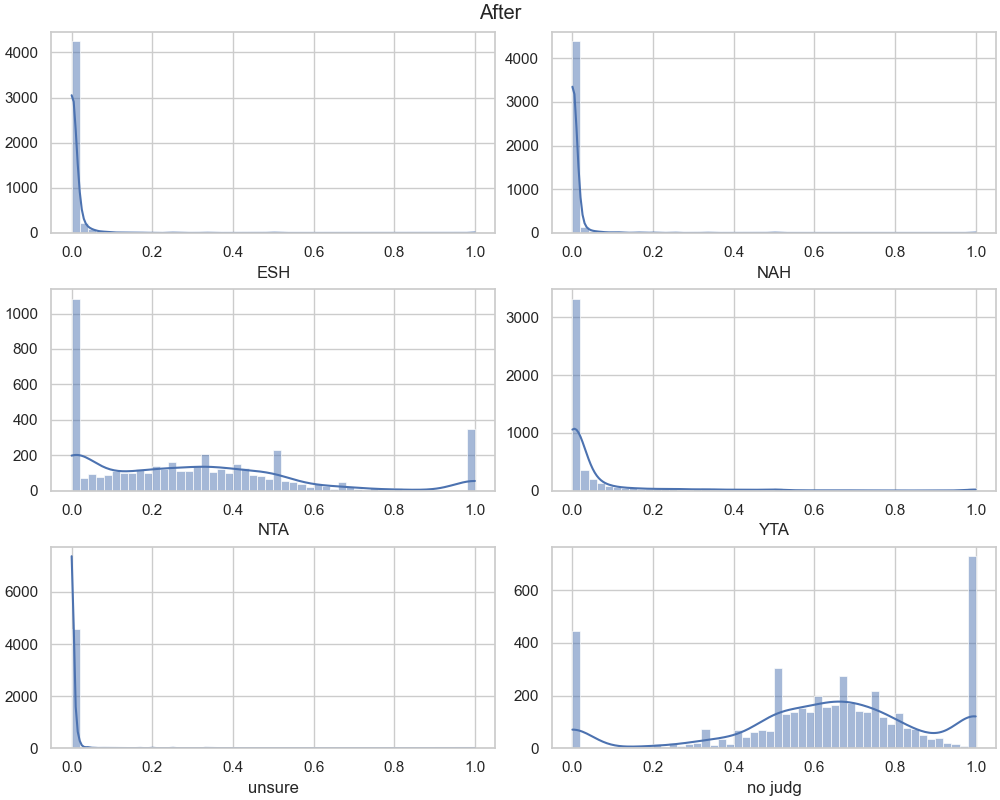}}
    
    \caption{Frequency of judgments before (a) and after (b) acknowledging the majority judgment, over all the 4,695 threads.}
    \label{fig:judg_distr_after}
\end{figure}

\clearpage
\subsection{Sentiment}\label{app:sentiment}
\begin{figure}[h!]
    \centering
    \includegraphics[width=0.5\textwidth, trim=0.1cm 0.7cm 0.1cm 2cm, clip]{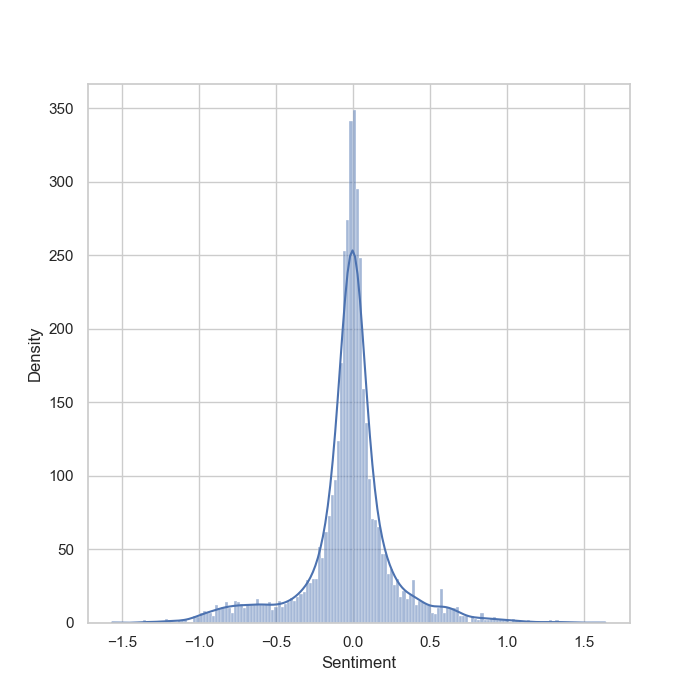} 
    \caption{Difference of threads' average sentiment before and after the majority judgment acknowledgment.}
    \label{fig:sentiment}
\end{figure}

We (i) extract the sentiment of each comment using VADER~\citep{Hutto_Gilbert_2014}, (ii) measure the average sentiment of a thread, (iii) compare distributions of sentiment before and after the eighteenth hour for all threads, and (iv) compute the difference between such distributions. No relevant difference was obtained, as shown in Figure~\ref{fig:sentiment}. The distribution is centered on zero, suggesting that discussions experience no change in sentiment.

\subsection{Polarization}\label{app:polarization}

For the plotted curves, $P_i(t)$ in Equation~\ref{eq:polarization} is averaged across all threads with the same final verdict:
\(
\bar{P}_v(t) = \frac{1}{N_v(t)} \sum_{i \in v} P_i(t)
\)
where \(v\) is a final-verdict group and \(N_v(t)\) is the number of threads in that group contributing at time \(t\).

\begin{figure}[h!]
    \centering
    \includegraphics[width=0.9\textwidth]{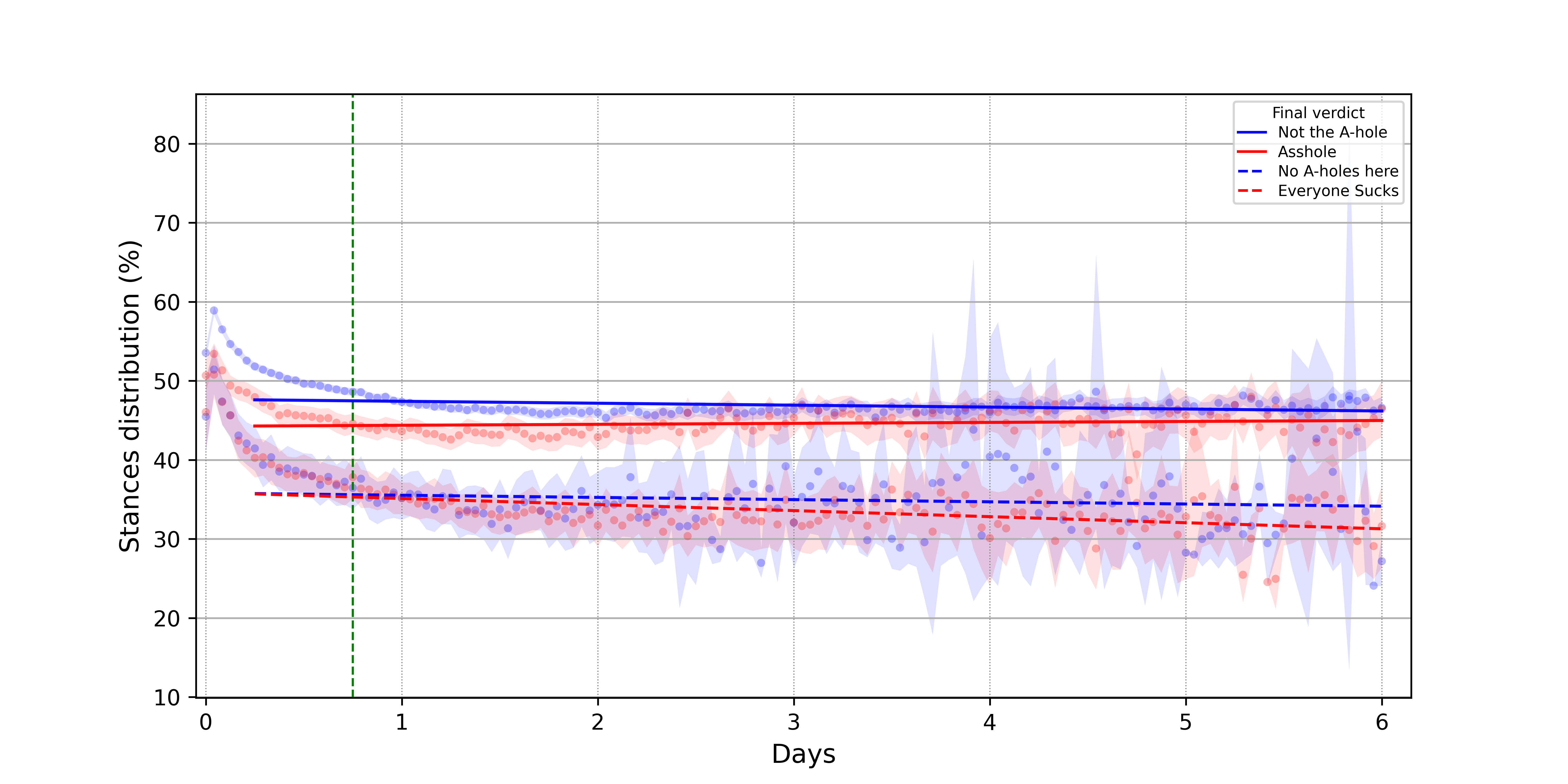}
    \caption{Polarization (distribution of stances) evolving over time, fitted over all discussions and grouped by final verdict. The dashed vertical line corresponds to the eighteenth hour, i.e., when the majority judgment is disclosed by the community and acknowledged by users.}
    \label{fig:polarization}
\end{figure}

\newpage
\subsection{Odd ratios}\label{app:odd_ratios}

Odd ratios are defined as:

\begin{equation}
    \text{OR}(p(d \mid C),\ p(d \mid \bar{C})) = 
\frac{p(d \mid C) \cdot (1 - p(d \mid \bar{C}))}
     {p(d \mid \bar{C}) \cdot (1 - p(d \mid C))}
\end{equation}

where:
\begin{itemize}
    \item $d$ is the variable representing each of the ten dimensions. It can take the value 1 ($d$) or 0 ($\bar{d})$, whether it appears or not in the comment.
    \item $C$ indicates conformity (individual judgment in agreement with the majority) and $\bar{C}$ indicates anti-conformity (individual judgment in disagreement with the majority).
    \item $p(d \mid C)$ is the probability of the dimension given a conformity behavior towards the majority.
    \item $p(d \mid \bar{C})$ is the probability of the dimension given an anti-conformity behavior towards the majority.
    \item $\frac{p (d\mid C)}{1 - p(d \mid C)}$ are the odds under agreement condition.
    \item $\frac{p (d\mid \bar{C})}{1 - p(d \mid \bar{C})}$ are the odds under disagreement condition.
\end{itemize}

We compute the OR first for all comments (independently of what is the majority judgments) and then distinguish by each of the four final verdicts (obtaining $OR_{ESH}$, $OR_{NAH}$, $OR_{NTA}$, $OR_{YTA}$).
The 95\% confidence intervals of the odds ratios are calculated as

\begin{equation}
    \text{CI} = z \cdot \sqrt{
\frac{1}{|d, C|} +
\frac{1}{|d, \bar{C}|} +
\frac{1}{|\bar{d}, C|} +
\frac{1}{|\bar{d}, \bar{C}|}
}
\end{equation}

where
$z=1.96$ is the critical value of the standard normal distribution and $|C\boldsymbol{\cdot},\boldsymbol{\cdot}|$ represents the cardinality of the set of comments with or without a given dimension ($d$ or $\bar{d}$) and with conformity or anti-conformity ($C$ or $\bar{C}$).

\begin{figure}[h]
    \centering
    \includegraphics[width=0.5\textwidth, trim=0.1cm 0.2cm 0.1cm 0.3cm, clip]{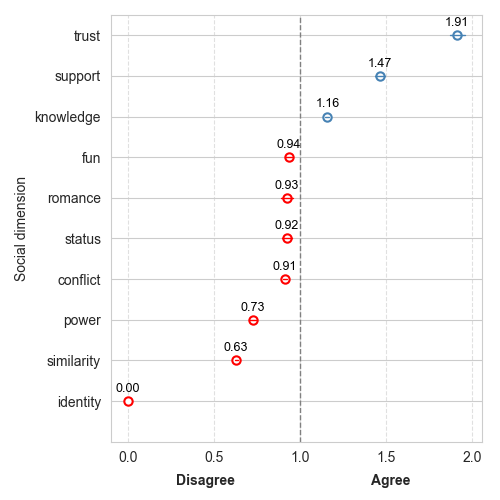}
    \caption{Odd ratios of the ten social dimensions aggregated for all the final verdicts.}
    \label{fig:odd_ratios_for_each_verdict}
\end{figure}

\newpage

\subsection{Local threshold analysis}\label{app:rdd}

\begin{table}[H]
\centering
\begin{tabular}{lrrrrrrrrr}
\hline
\textbf{Judgment}  & \textbf{$\hat{\tau}$} & \textbf{SE} & \textbf{CI$_{95\%}$ Low} & \textbf{CI$_{95\%}$ High} & \textbf{$p$-value} & \textbf{$n$} & \textbf{$\beta_1$} & \textbf{$\beta_1 + \beta_2$} \\
\hline
\multicolumn{9}{c}{Verdict: \textcolor{red}{YTA}} \\
\hline
no\_vote & -0.74 & 0.06 & -0.87 & -0.62 & 2.32e-30 & 699109 & 0.26 & 0.18 \\
unsure   & -0.04 & 0.00 & -0.05 & -0.04 & 8.16e-49 & 699109 & 0.00 & 0.00 \\
YTA      & 1.26  & 0.09 & 1.08  & 1.45  & 3.93e-41 & 699109 & -0.32 & -0.22 \\
ESH      & 0.01  & 0.01 & -0.01 & 0.02  & 5.79e-01 & 699109 & -0.00 & 0.00 \\
NTA      & -0.41 & 0.03 & -0.48 & -0.35 & 1.28e-33 & 699109 & 0.05 & 0.04 \\
NAH      & -0.07 & 0.01 & -0.08 & -0.05 & 2.88e-19 & 699109 & -0.00 & -0.01 \\
\hline
\multicolumn{9}{c}{Verdict: \textcolor{red}{ESH}} \\
\hline
no\_vote & 0.85  & 0.12 & 0.61  & 1.08  & 3.81e-12 & 114286 & 0.01 & 0.16 \\
unsure   & -0.10 & 0.01 & -0.11 & -0.08 & 2.76e-24 & 114286 & -0.02 & -0.00 \\
YTA      & -0.20 & 0.08 & -0.35 & -0.05 & 8.60e-03 & 114286 & 0.06 & -0.08 \\
ESH      & -0.42 & 0.07 & -0.55 & -0.29 & 4.39e-10 & 114286 & -0.03 & -0.06 \\
NTA      & -0.27 & 0.11 & -0.48 & -0.05 & 1.43e-02 & 114286 & -0.01 & -0.00 \\
NAH      & 0.14  & 0.02 & 0.10  & 0.18  & 1.96e-12 & 114286 & 0.00 & -0.01 \\
\hline
\multicolumn{9}{c}{Verdict: \textcolor{blue}{NTA}} \\
\hline
no\_vote & -4.68 & 0.28 & -5.23 & -4.14 & 5.82e-63 & 60166 & 0.18 & 0.41 \\
unsure   & -0.08 & 0.01 & -0.10 & -0.07 & 3.77e-21 & 60166 & -0.01 & 0.00 \\
YTA      & -0.23 & 0.11 & -0.45 & -0.01 & 4.26e-02 & 60166 & -0.21 & 0.19 \\
ESH      & -0.04 & 0.02 & -0.08 & -0.00 & 4.37e-02 & 60166 & -0.01 & 0.00 \\
NTA      & 4.52  & 0.25 & 4.03  & 5.02  & 2.18e-72 & 60166 & 0.12 & -0.45 \\
NAH      & 0.51  & 0.11 & 0.30  & 0.73  & 2.69e-06 & 60166 & -0.07 & -0.16 \\
\hline
\multicolumn{9}{c}{Verdict: \textcolor{blue}{NAH}} \\
\hline
no\_vote & -0.36 & 0.04 & -0.43 & -0.29 & 9.03e-23 & 2148633 & 0.23 & 0.11 \\
unsure   & -0.04 & 0.00 & -0.04 & -0.04 & 2.01e-132 & 2148633 & 0.00 & -0.00 \\
YTA      & -0.23 & 0.01 & -0.26 & -0.21 & 6.15e-68 & 2148633 & 0.08 & -0.02 \\
ESH      & -0.06 & 0.01 & -0.07 & -0.04 & 6.47e-17 & 2148633 & -0.01 & -0.01 \\
NTA      & 0.75  & 0.05 & 0.65  & 0.84  & 1.09e-54 & 2148633 & -0.31 & -0.06 \\
NAH      & -0.06 & 0.00 & -0.07 & -0.05 & 6.03e-41 & 2148633 & 0.01 & -0.01 \\
\hline
\end{tabular}
\caption{Regression discontinuities estimate at the 18-hour verdict. $\hat{\tau}$ is the estimated jump at the threshold 
, while SE and CIs represent the uncertainty around the estimate. $\hat{\tau}$ is expressed in percentage points (pp) and the slopes are in percentage points per hour. The total number of data points that fall within the $\pm$12-hour bandwidth used to estimate the discontinuity is 3{,}022{,}194.}\label{tab:rdd}
\end{table}

\newpage

\begin{figure}[H]
\centering
\footnotesize

\subfloat[Judgment expressed: \textcolor{red}{YTA}]{
\includegraphics[width=0.8\textwidth]{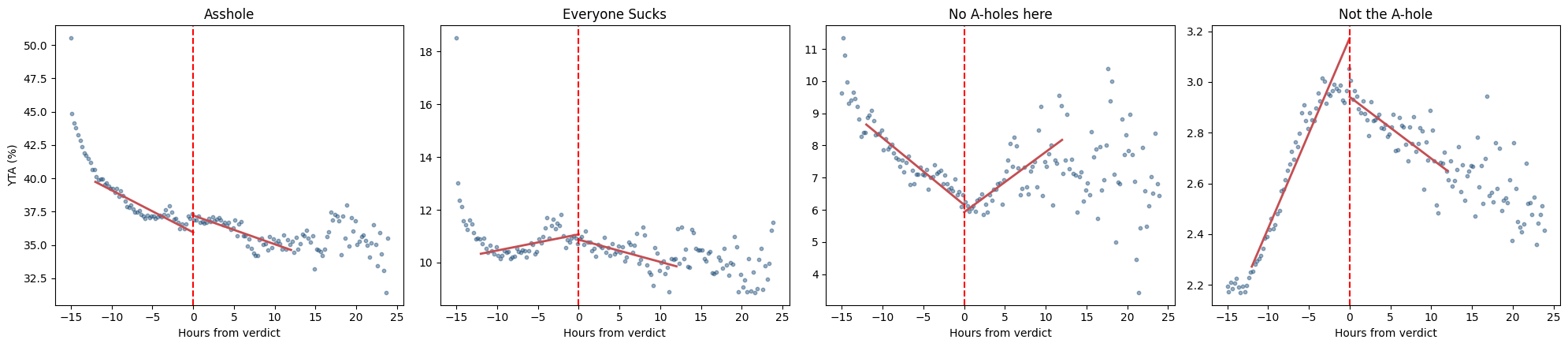}}

\subfloat[Judgment expressed: \textcolor{red}{ESH}]{\includegraphics[width=0.8\textwidth]{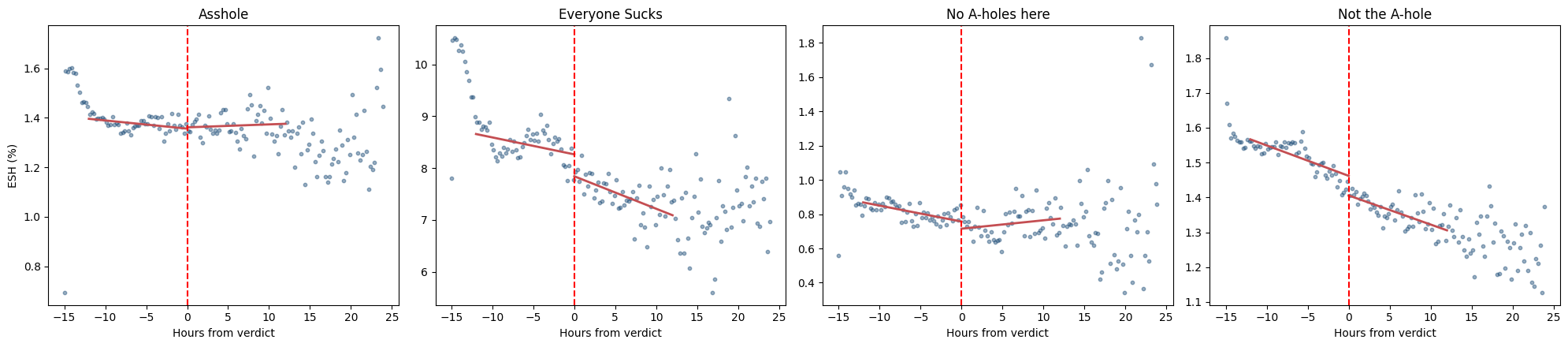}}

\subfloat[Judgment expressed: \textcolor{blue}{NAH}]{\includegraphics[width=0.8\textwidth]{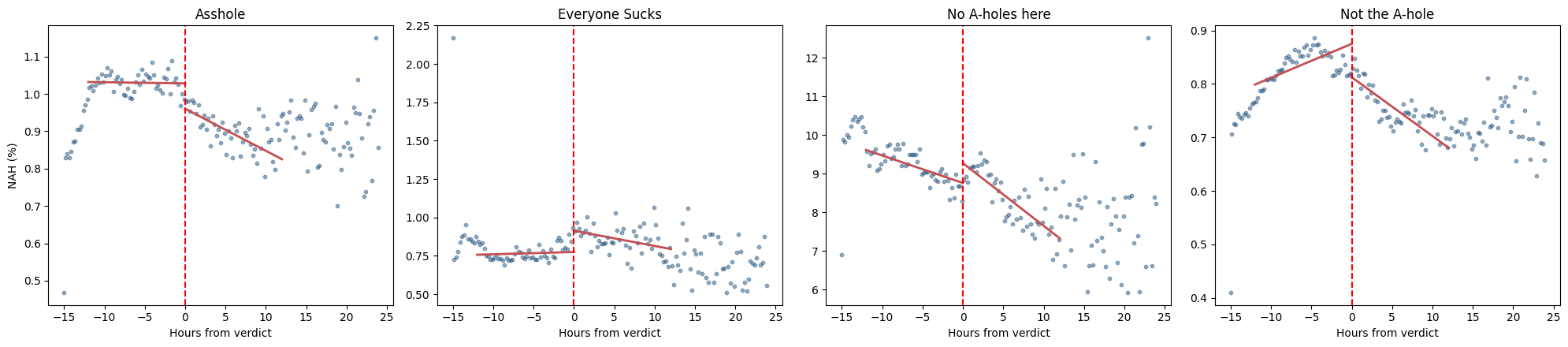}}

\subfloat[Judgment expressed: \textcolor{blue}{NTA}]{\includegraphics[width=0.8\textwidth]{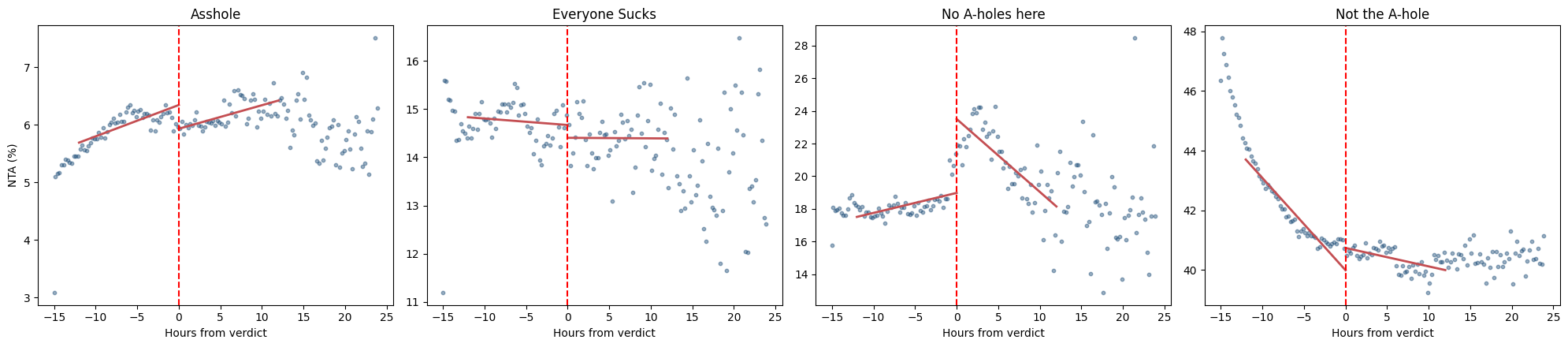}}

\subfloat[Judgment expressed: unsure]{\includegraphics[width=0.8\textwidth]{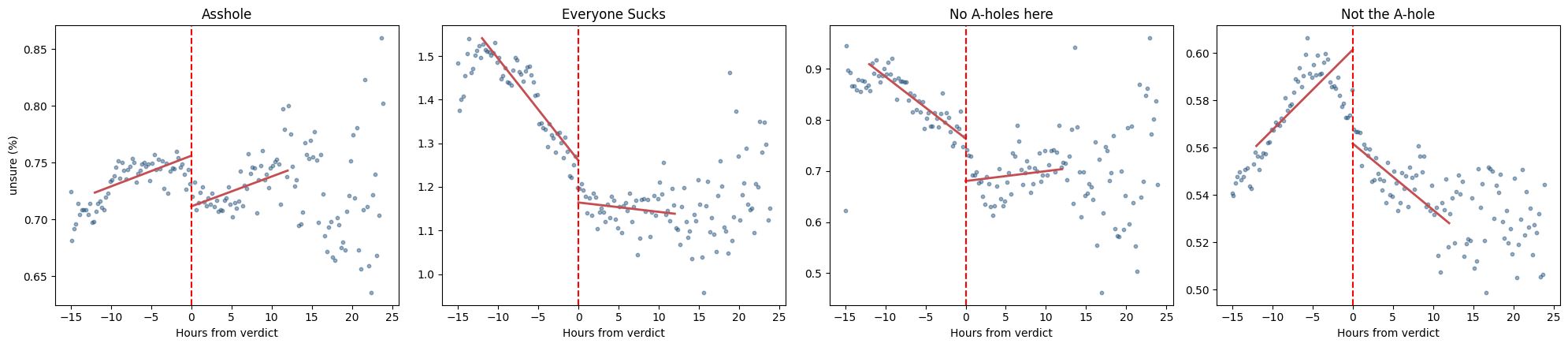}}

\subfloat[No judgment expressed]{\includegraphics[width=0.8\textwidth]{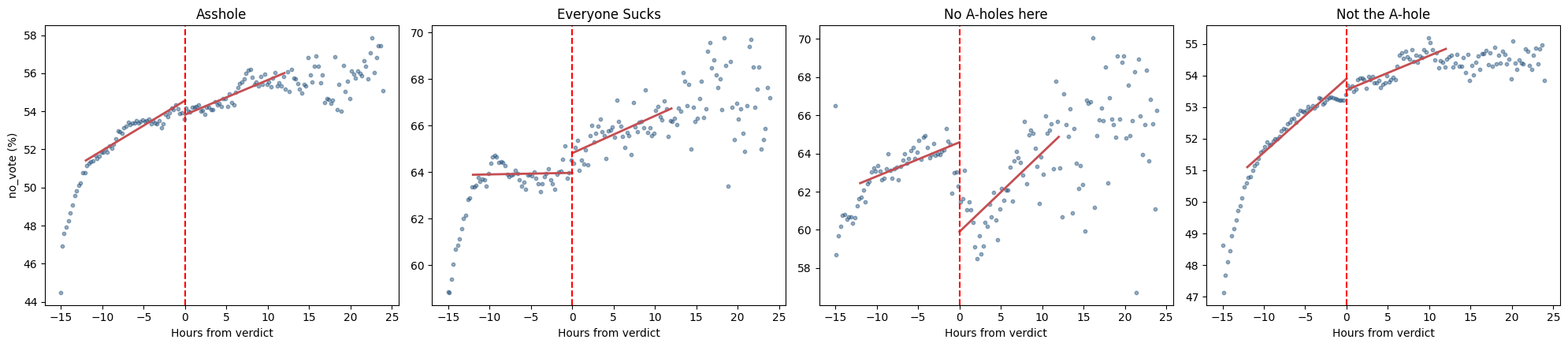}}

\caption{Results of the local threshold analysis for every verdict. The intercept of the post-verdict regression line indicates the discontinuity at the threshold ($\alpha + \hat{\tau}$). The slope of the post-verdict regression line ($\beta_1 + \beta_2$) reveals the trend of judgment expressions after the threshold. Cumulative averages are reported every 30 minutes for a better readability. The $y$ axis represents the average running count of every judgment, across all threads, that fall in the corresponding $x$ timestamp.}\label{fig:rdd_all}
\end{figure}

\newpage
\subsection{Bayesian models}\label{app:other_models}

\algrenewcommand\algorithmicrequire{\textbf{Input:}}
\algrenewcommand\algorithmicensure{\textbf{Output:}}
\begin{algorithm}[h]
\footnotesize
\begin{algorithmic}[1]
\Require Data for individual judgments \textit{before} the verdict: $B_{ESH}, B_{NAH}, B_{NTA}, B_{YTA}, B_{u}, B_{nj}$
\Require Data for individual judgments \textit{after} the verdict: $A_{ESH}, A_{NAH}, A_{NTA}, A_{YTA}, A_{u}, A_{nj}$
\Require Mean values for each individual judgment: $\overline{ESH}, \overline{NAH}, \overline{NTA}, \overline{YTA}, \overline{u}, \overline{nj}$
\Require List of final verdicts for each thread: $V$
\Require List of possible verdicts: $possible\_v = [ESH, NAH, NTA, YTA]$
\Ensure A list of posterior distributions for each model

\Statex
\State $\textit{vars} \gets [[B_{ESH}, A_{ESH}], [B_{NAH}, A_{NAH}], [B_{NTA}, A_{NTA}], [B_{YTA}, A_{YTA}], [B_{u}, A_{u}], [B_{nj}, A_{nj}]]$
\State $\textit{means} \gets [\overline{ESH}, \overline{NAH}, \overline{NTA}, \overline{YTA}, \overline{u}, \overline{nj}]$
\State $\textit{posteriors} \gets []$

\For{$\textit{var\_idx}$ from $0$ to $\text{length}(\textit{vars}) - 1$}
    \State \textbf{Initialize} probabilistic model $m$
    \State \textbf{Define} priors
    \Statex \quad \quad $\textit{prior\_mean} \gets \text{mean}(\textit{vars}[\textit{var\_idx}][0])$
    \Statex \quad \quad $\textit{prior\_sd} \gets \text{std\_dev}(\textit{vars}[\textit{var\_idx}][0])$
    \Statex \quad \quad $\alpha \sim \text{Normal}(\textit{prior\_mean}, \textit{prior\_sd}, \text{shape}=\text{length}(\textit{possible\_v}))$
    \Statex \quad \quad $\sigma \sim \text{Uniform}(0, 1)$
    \Statex \quad \quad $\beta \sim \text{Normal}(0, 10, \text{shape}=\text{length}(\textit{possible\_v}))$
    \State \textbf{Define} the likelihood
    \Statex \quad \quad $\mu = \alpha[V] + \beta[V] (\textit{vars}[\textit{var\_idx}][0] - \textit{means}[\textit{var\_idx}])$
    \Statex \quad \quad $Y_{likelihood} \sim \text{Normal}(\mu, \sigma, \text{observed}=\textit{vars}[\textit{var\_idx}][1])$
    \State \textbf{Sample} the posterior 
    \Statex \quad \quad $p \gets$ perform a MCMC sampling drawing from the posterior 
    \Statex \quad \quad $\textit{posteriors}.\text{append}(p)$
    \State \textbf{Output} $p$ statistics with $0.89$ HDI probability
\EndFor
\State \textbf{Return} $\textit{posteriors}$
\end{algorithmic}
\captionsetup{font=footnotesize}
\caption{Implementation of the Bayesian multivariate regression. We use the probabilistic programming library for Python (\texttt{PyMC}; \url{https://www.pymc.io/welcome.html}). The MCMC sample used is NUTS, a highly efficient and robust Hamiltonian Monte Carlo (HMC) algorithm for automatic tuning of the parameters. The corresponding desired average acceptance probability is set to 0.95, which guarantees a precise exploration of the posterior.}
\label{alg1}
\end{algorithm}

To confirm the robustness of our results, we tried alternative Bayesian models.
The first two handle the final verdict variable $V$ differently, the third differentiates the $\beta$ variable in $\beta_1$ and $\beta_2$, and the fourth assumes variables to be non-identically independently distributed. All these models show a good performance (good convergence, negligible noise, and considerable sample size), but their different structure makes them not the best possible models to answer our \textbf{RQ}.

\newpage
\subsubsection{Stratification of the final verdict}

In the model represented by Equation \ref{eq:4alpha_1beta}, we present a fixed effects model with a common slope $\beta$.
$\alpha$ is again a verdict-specific intercept, capturing the shifts dependent on the categorical variable $v$ (i.e., each $V[v]$ has its own baseline mean). However, the effect of $B_j$ on $\mu_i$ is assumed constant across all majority judgments $V[v]$, meaning that the steepness of the regression slope is the same for all verdicts. This is an oversimplifying conjecture: we are assuming that the strength (or direction) of the relationship between individual judgments and the predicted mean is consistent across all groups. As a consequence, results (Table~\ref{tab:4alpha_1beta}) show such limitation.
$\alpha$ parameters are different for each $V[v]$ within each of the six models: by forcing a single $\beta$, we assume that the effect of $(B_{j}-\bar{B})$ is identical across verdicts, even though the verdicts start from different points. This approach prevents the detection and quantification of verdict-specific differences in individual judgments.

\begin{equation}\label{eq:4alpha_1beta}
    \mu_i = \alpha_{V[v]} + \beta (B_{j}-\bar{B}) \quad \forall j \in J
\end{equation}

\begin{table}[h!]
\captionsetup{font=footnotesize}
\centering
\footnotesize
\scalebox{0.9}{\begin{tabular}{lrrrrrrrrr}

\multicolumn{10}{l}{\textbf{Model 1) Judgment expressed: \textcolor{red}{YTA}}} \\
\midrule
\textbf{Param} & \textbf{mean} & \textbf{sd} & \textbf{hdi\_5.5\%} & \textbf{hdi\_94.5\%} & \textbf{mcse\_mean} & \textbf{mcse\_sd} & \textbf{ess\_bulk} & \textbf{ess\_tail} & \textbf{r\_hat} \\
\midrule
\textcolor{red}{$\alpha_{YTA}$} & \textbf{0.369} & 0.020 & 0.338 & 0.402 & 0.0 & 0.0 & 6102.0  & 6307.0 & 1.0 \\
\textcolor{red}{$\alpha_{ESH}$} & 0.180 & 0.046 & 0.105 & 0.253 & 0.0 & 0.0 & 9868.0  & 6586.0 & 1.0 \\
\textcolor{blue}{$\alpha_{NAH}$} & 0.126 & 0.044 & 0.052 & 0.192 & 0.0 & 0.0 & 9210.0  & 6421.0 & 1.0 \\
\textcolor{blue}{$\alpha_{NTA}$} & 0.035 & 0.006 & 0.026 & 0.045 & 0.0 & 0.0 & 7384.0  & 6738.0 & 1.0 \\
$\beta$    & 0.188 & 0.036 & 0.133 & 0.247 & 0.0 & 0.0 & 5640.0  & 5331.0 & 1.0 \\
$\sigma$    & 0.157 & 0.004 & 0.151 & 0.163 & 0.0 & 0.0 & 8648.0  & 6502.0 & 1.0 \\
\bottomrule\\[0.02ex]

\multicolumn{10}{l}{\textbf{Model 2) Judgment expressed: \textcolor{red}{ESH}}} \\
\midrule
\textbf{Param} & \textbf{mean} & \textbf{sd} & \textbf{hdi\_5.5\%} & \textbf{hdi\_94.5\%} & \textbf{mcse\_mean} & \textbf{mcse\_sd} & \textbf{ess\_bulk} & \textbf{ess\_tail} & \textbf{r\_hat} \\
\midrule
\textcolor{red}{$\alpha_{YTA}$} & 0.022 & 0.009 & 0.009 & 0.037 & 0.0 & 0.0 & 12025.0 & 5892.0 & 1.0 \\
\textcolor{red}{$\alpha_{ESH}$} & 0.058 & 0.028 & 0.016 & 0.102 & 0.0 & 0.0 & 12491.0 & 7074.0 & 1.0 \\
\textcolor{blue}{$\alpha_{NAH}$} & 0.002 & 0.025 & -0.037 & 0.044 & 0.0 & 0.0 & 12752.0 & 6515.0 & 1.0 \\
\textcolor{blue}{$\alpha_{NTA}$} & 0.014 & 0.003 & 0.009 & 0.019 & 0.0 & 0.0 & 11422.0 & 6309.0 & 1.0 \\
$\beta$    & 0.120 & 0.037 & 0.062 & 0.179 & 0.0 & 0.0 & 11507.0 & 6916.0 & 1.0 \\
$\sigma$    & 0.091 & 0.002 & 0.087 & 0.094 & 0.0 & 0.0 & 9119.0  & 6395.0 & 1.0 \\
\bottomrule\\[0.02ex]

\multicolumn{10}{l}{\textbf{Model 3) Judgment expressed: \textcolor{blue}{NAH}}} \\
\midrule
\textbf{Param} & \textbf{mean} & \textbf{sd} & \textbf{hdi\_5.5\%} & \textbf{hdi\_94.5\%} & \textbf{mcse\_mean} & \textbf{mcse\_sd} & \textbf{ess\_bulk} & \textbf{ess\_tail} & \textbf{r\_hat} \\
\midrule
\textcolor{red}{$\alpha_{YTA}$} & 0.017 & 0.007 & 0.005 & 0.028 & 0.0 & 0.0 & 10531.0 & 6301.0 & 1.0 \\
\textcolor{red}{$\alpha_{ESH}$} & 0.006 & 0.023 & -0.031 & 0.042 & 0.0 & 0.0 & 10724.0 & 6612.0 & 1.0 \\
\textcolor{blue}{$\alpha_{NAH}$} & \textbf{0.224} & 0.022 & 0.189 & 0.258 & 0.0 & 0.0 & 9912.0  & 6770.0 & 1.0 \\
\textcolor{blue}{$\alpha_{NTA}$} & 0.008 & 0.003 & 0.003 & 0.012 & 0.0 & 0.0 & 9688.0  & 6628.0 & 1.0 \\
$\beta$    & 0.153 & 0.036 & 0.096 & 0.210 & 0.0 & 0.0 & 11276.0 & 6519.0 & 1.0 \\
$\sigma$    & 0.076 & 0.002 & 0.073 & 0.079 & 0.0 & 0.0 & 12751.0 & 6680.0 & 1.0 \\
\bottomrule\\[0.02ex]

\multicolumn{10}{l}{\textbf{Model 4) Judgment expressed: \textcolor{blue}{NTA}}} \\
\midrule
\textbf{Param} & \textbf{mean} & \textbf{sd} & \textbf{hdi\_5.5\%} & \textbf{hdi\_94.5\%} & \textbf{mcse\_mean} & \textbf{mcse\_sd} & \textbf{ess\_bulk} & \textbf{ess\_tail} & \textbf{r\_hat} \\
\midrule
\textcolor{red}{$\alpha_{YTA}$} & 0.207 & 0.030 & 0.178 & 0.274 & 0.0 & 0.0 & 7519.0  & 6965.0 & 1.0 \\
\textcolor{red}{$\alpha_{ESH}$} & 0.107 & 0.082 & 0.175 & 0.434 & 0.001 & 0.001 & 9238.0 & 6544.0 & 1.0 \\
\textcolor{blue}{$\alpha_{NAH}$} & \textbf{0.484} & 0.074 & 0.369 & 0.606 & 0.001 & 0.001 & 9528.0 & 6392.0 & 1.0 \\
\textcolor{blue}{$\alpha_{NTA}$} & \textbf{0.557} & 0.010 & 0.541 & 0.574 & 0.0 & 0.0 & 11399.0 & 5967.0 & 1.0 \\
$\beta$    & 0.298 & 0.034 & 0.244 & 0.351 & 0.0 & 0.0 & 6275.0  & 6955.0 & 1.0 \\
$\sigma$    & 0.270 & 0.007 & 0.260 & 0.281 & 0.0 & 0.0 & 8143.0  & 6681.0 & 1.0 \\
\bottomrule\\[0.02ex]

\multicolumn{10}{l}{\textbf{Model 5) Judgment expressed: unsure}} \\
\midrule
\textbf{Param} & \textbf{mean} & \textbf{sd} & \textbf{hdi\_5.5\%} & \textbf{hdi\_94.5\%} & \textbf{mcse\_mean} & \textbf{mcse\_sd} & \textbf{ess\_bulk} & \textbf{ess\_tail} & \textbf{r\_hat} \\
\midrule
\textcolor{red}{$\alpha_{YTA}$} & 0.004 & 0.002 & 0.001 & 0.008 & 0.0 & 0.0 & 11530.0 & 6159.0 & 1.0 \\
\textcolor{red}{$\alpha_{ESH}$} & 0.006 & 0.007 & -0.005 & 0.017 & 0.0 & 0.0 & 9794.0  & 6607.0 & 1.0 \\
\textcolor{blue}{$\alpha_{NAH}$} & 0.001 & 0.006 & -0.009 & 0.011 & 0.0 & 0.0 & 13309.0 & 6654.0 & 1.0 \\
\textcolor{blue}{$\alpha_{NTA}$} & 0.004 & 0.001 & 0.003 & 0.006 & 0.0 & 0.0 & 12941.0 & 6137.0 & 1.0 \\
$\beta$    & 0.041 & 0.033 & -0.013 & 0.093 & 0.0 & 0.0 & 14274.0 & 6560.0 & 1.0 \\
$\sigma$    & 0.023 & 0.001 & 0.022 & 0.024 & 0.0 & 0.0 & 12270.0 & 6182.0 & 1.0 \\
\bottomrule\\[0.02ex]

\multicolumn{10}{l}{\textbf{Model 6) No judgment expressed}} \\
\midrule
\textbf{Param} & \textbf{mean} & \textbf{sd} & \textbf{hdi\_5.5\%} & \textbf{hdi\_94.5\%} & \textbf{mcse\_mean} & \textbf{mcse\_sd} & \textbf{ess\_bulk} & \textbf{ess\_tail} & \textbf{r\_hat} \\
\midrule
\textcolor{red}{$\alpha_{YTA}$} & 0.402 & 0.025 & 0.361 & 0.443 & 0.0 & 0.0 & 12672.0 & 6592.0 & 1.0 \\
\textcolor{red}{$\alpha_{ESH}$} & 0.472 & 0.078 & 0.350 & 0.597 & 0.001 & 0.0 & 15213.0 & 6707.0 & 1.0 \\
\textcolor{blue}{$\alpha_{NAH}$} & 0.187 & 0.072 & 0.071 & 0.298 & 0.001 & 0.001 & 9940.0  & 5873.0 & 1.0 \\
\textcolor{blue}{$\alpha_{NTA}$} & 0.374 & 0.010 & 0.359 & 0.390 & 0.0 & 0.0 & 11594.0 & 6552.0 & 1.0 \\
$\beta$    & 0.299 & 0.035 & 0.242 & 0.353 & 0.0 & 0.0 & 8336.0  & 6518.0 & 1.0 \\
$\sigma$    & 0.258 & 0.007 & 0.247 & 0.268 & 0.0 & 0.0 & 7569.0  & 6376.0 & 1.0 \\
\bottomrule\\[0.02ex]

\end{tabular}}
\caption{Fixed effect model with common $\beta$ parameter.}\label{tab:4alpha_1beta}
\end{table}

In the model represented by Equation \ref{eq:1alpha_4beta}, different verdicts do not have different starting points but only different slopes. The effect of the deviation of individual judgments from the mean is modulated by each different verdict (as in Section \ref{subsec:model}), but $\alpha$ is now a global intercept. This means there is one common baseline mean ($\mathbb{E}[\mu_i]$ when $B_{j}-\bar{B} =0$) for all observations, regardless of the value of $V[v]$. The assumption, in this case, is again oversimplifying: we expect the verdict not to affect the baseline of $\mu_i$, which implies that the initial average value does not change depending on the different majority judgments obtained (see Table~\ref{tab:1alpha_4beta}). 
By allowing the baseline to vary with $V[v]$, as in Equation \ref{eq:1}, we account for unmeasured differences that might influence the baseline (confounding factors). This is especially important in observational studies, where ignoring baseline differences can lead to biased estimates of other effects in the model.

\begin{equation}\label{eq:1alpha_4beta}
    \mu_i = \alpha + \beta (B_{j}-\bar{B})V[v] \quad \forall j \in J
\end{equation}

\begin{table}[h!]
\captionsetup{font=footnotesize}
\centering
\footnotesize

\scalebox{0.9}{\begin{tabular}{lrrrrrrrrr}

\multicolumn{10}{l}{\textbf{Model 1) Judgment expressed: \textcolor{red}{YTA}}} \\
\midrule
\textbf{Param} & \textbf{mean} & \textbf{sd} & \textbf{hdi\_5.5\%} & \textbf{hdi\_94.5\%} & \textbf{mcse\_mean} & \textbf{mcse\_sd} & \textbf{ess\_bulk} & \textbf{ess\_tail} & \textbf{r\_hat} \\
\midrule
alpha & 0.502 & 0.128 & 0.302 & 0.710 & 0.002 & 0.001 & 4063.0 & 5343.0 & 1.0 \\
\textcolor{red}{$\beta_{YTA}$} & 0.826 & 0.091 & 0.680 & 0.972 & 0.001 & 0.001 & 4912.0 & 5621.0 & 1.0 \\
\textcolor{red}{$\beta_{ESH}$} & 1.457 & 0.421 & 0.771 & 2.108 & 0.004 & 0.003 & 8832.0 & 5849.0 & 1.0 \\
\textcolor{blue}{$\beta_{NAH}$} & 0.604 & 0.357 & 0.061 & 1.204 & 0.004 & 0.003 & 7946.0 & 6662.0 & 1.0 \\
\textcolor{blue}{$\beta_{NTA}$} & 0.989 & 0.167 & 0.732 & 1.264 & 0.002 & 0.002 & 4724.0 & 6036.0 & 1.0 \\
sigma & 0.987 & 0.012 & 0.971 & 1.000 & 0.000 & 0.000 & 6344.0 & 3467.0 & 1.0 \\
\bottomrule\\[0.02ex]

\multicolumn{10}{l}{\textbf{Model 2) Judgment expressed: \textcolor{red}{ESH}}} \\
\midrule
\textbf{Param} & \textbf{mean} & \textbf{sd} & \textbf{hdi\_5.5\%} & \textbf{hdi\_94.5\%} & \textbf{mcse\_mean} & \textbf{mcse\_sd} & \textbf{ess\_bulk} & \textbf{ess\_tail} & \textbf{r\_hat} \\
\midrule
alpha & 0.003 & 0.064 & -0.101 & 0.106 & 0.001 & 0.001 & 7951.0 & 6664.0 & 1.0 \\
\textcolor{red}{$\beta_{YTA}$} & 0.294 & 0.112 & 0.112  & 0.476 & 0.001 & 0.001 & 8945.0 & 6508.0 & 1.0 \\
\textcolor{red}{$\beta_{ESH}$} & 0.394 & 0.110 & 0.210  & 0.562 & 0.001 & 0.001 & 8263.0 & 6174.0 & 1.0 \\
\textcolor{blue}{$\beta_{NAH}$} & 0.020 & 0.334 & -0.502 & 0.556 & 0.003 & 0.003 & 9548.0 & 7165.0 & 1.0 \\
\textcolor{blue}{$\beta_{NTA}$} & 0.168 & 0.144 & -0.052 & 0.401 & 0.002 & 0.001 & 7911.0 & 6789.0 & 1.0 \\
sigma & 0.991 & 0.008 & 0.981  & 1.000 & 0.000 & 0.000 & 6497.0 & 3881.0 & 1.0 \\
\bottomrule\\[0.02ex]

\multicolumn{10}{l}{\textbf{Model 3) Judgment expressed: \textcolor{blue}{NAH}}} \\
\midrule
\textbf{Param} & \textbf{mean} & \textbf{sd} & \textbf{hdi\_5.5\%} & \textbf{hdi\_94.5\%} & \textbf{mcse\_mean} & \textbf{mcse\_sd} & \textbf{ess\_bulk} & \textbf{ess\_tail} & \textbf{r\_hat} \\
\midrule
alpha & 0.182 & 0.069 & 0.073 & 0.294 & 0.001 & 0.001 & 8703.0 & 6893.0 & 1.0 \\
\textcolor{red}{$\beta_{YTA}$} & 0.251 & 0.084 & 0.115 & 0.383 & 0.001 & 0.001 & 8898.0 & 6464.0 & 1.0 \\
\textcolor{red}{$\beta_{ESH}$} & 0.367 & 0.526 & -0.459 & 1.212 & 0.005 & 0.005 & 9201.0 & 6800.0 & 1.0 \\
\textcolor{blue}{$\beta_{NAH}$} & 0.985 & 0.065 & 0.879 & 1.087 & 0.001 & 0.000 & 9039.0 & 6311.0 & 1.0 \\
\textcolor{blue}{$\beta_{NTA}$} & 1.078 & 0.152 & 0.841 & 1.326 & 0.002 & 0.001 & 9305.0 & 6976.0 & 1.0 \\
sigma & 0.997 & 0.003 & 0.994 & 1.000 & 0.000 & 0.000 & 6989.0 & 4458.0 & 1.0 \\
\bottomrule\\[0.02ex]

\multicolumn{10}{l}{\textbf{Model 4) Judgment expressed: \textcolor{blue}{NTA}}} \\
\midrule
\textbf{Param} & \textbf{mean} & \textbf{sd} & \textbf{hdi\_5.5\%} & \textbf{hdi\_94.5\%} & \textbf{mcse\_mean} & \textbf{mcse\_sd} & \textbf{ess\_bulk} & \textbf{ess\_tail} & \textbf{r\_hat} \\
\midrule
alpha & 0.805 & 0.109 & 0.636 & 0.983 & 0.002 & 0.001 & 4139.0 & 5579.0 & 1.0 \\
\textcolor{red}{$\beta_{YTA}$} & 1.026 & 0.114 & 0.847 & 1.209 & 0.002 & 0.001 & 4227.0 & 5148.0 & 1.0 \\
\textcolor{red}{$\beta_{ESH}$} & 1.058 & 0.384 & 0.436 & 1.656 & 0.004 & 0.003 & 7343.0 & 6238.0 & 1.0 \\
\textcolor{blue}{$\beta_{NAH}$} & 1.208 & 0.380 & 0.625 & 1.843 & 0.005 & 0.003 & 7116.0 & 5788.0 & 1.0 \\
\textcolor{blue}{$\beta_{NTA}$} & 0.680 & 0.098 & 0.520 & 0.831 & 0.001 & 0.001 & 4703.0 & 5567.0 & 1.0 \\
sigma & 0.935 & 0.036 & 0.886 & 0.997 & 0.000 & 0.000 & 4514.0 & 2723.0 & 1.0 \\
\bottomrule\\[0.02ex]

\multicolumn{10}{l}{\textbf{Model 5) Judgment expressed: unsure}} \\
\midrule
\textbf{Param} & \textbf{mean} & \textbf{sd} & \textbf{hdi\_5.5\%} & \textbf{hdi\_94.5\%} & \textbf{mcse\_mean} & \textbf{mcse\_sd} & \textbf{ess\_bulk} & \textbf{ess\_tail} & \textbf{r\_hat} \\
\midrule
alpha & -0.029 & 0.037 & -0.089 & 0.029 & 0.000 & 0.000 & 11615.0 & 6697.0 & 1.0 \\
\textcolor{red}{$\beta_{YTA}$}  & 0.006  & 0.034 & -0.049 & 0.059 & 0.000 & 0.000 & 11483.0 & 7761.0 & 1.0 \\
\textcolor{red}{$\beta_{ESH}$} & 0.018  & 0.098 & -0.137 & 0.174 & 0.001 & 0.001 & 12253.0 & 7661.0 & 1.0 \\
\textcolor{blue}{$\beta_{NAH}$} & 0.002  & 0.208 & -0.341 & 0.323 & 0.002 & 0.002 & 15770.0 & 7327.0 & 1.0 \\
\textcolor{blue}{$\beta_{NTA}$} & -0.024 & 0.051 & -0.103 & 0.062 & 0.000 & 0.000 & 11808.0 & 7668.0 & 1.0 \\
sigma & 0.576  & 0.026 & 0.534  & 0.617 & 0.000 & 0.000 & 16069.0 & 7300.0 & 1.0 \\
\bottomrule\\[0.02ex]

\multicolumn{10}{l}{\textbf{Model 6) No judgment expressed}} \\
\midrule
\textbf{Param} & \textbf{mean} & \textbf{sd} & \textbf{hdi\_5.5\%} & \textbf{hdi\_94.5\%} & \textbf{mcse\_mean} & \textbf{mcse\_sd} & \textbf{ess\_bulk} & \textbf{ess\_tail} & \textbf{r\_hat} \\
\midrule
alpha & -0.988 & 0.057 & -1.081 & -0.899 & 0.001 & 0.000 & 7996.0 & 5646.0 & 1.0 \\
\textcolor{red}{$\beta_{YTA}$} & 0.518  & 0.083 & 0.384  & 0.648  & 0.001 & 0.001 & 8247.0 & 6169.0 & 1.0 \\
\textcolor{red}{$\beta_{ESH}$} & 1.070  & 0.331 & 0.561  & 1.612  & 0.003 & 0.003 & 8971.0 & 6572.0 & 1.0 \\
\textcolor{blue}{$\beta_{NAH}$} & 0.326  & 0.270 & -0.108 & 0.740  & 0.003 & 0.003 & 6950.0 & 5737.0 & 1.0 \\
\textcolor{blue}{$\beta_{NTA}$} & 0.505  & 0.089 & 0.357  & 0.640  & 0.001 & 0.001 & 8666.0 & 6344.0 & 1.0 \\
sigma & 0.907  & 0.039 & 0.844  & 0.970  & 0.001 & 0.000 & 4956.0 & 2883.0 & 1.0 \\
\hline\hline\\
\end{tabular}}
\caption{Model with common baseline.}\label{tab:1alpha_4beta}
\end{table}

\newpage
\subsubsection{Non-centered individual judgments}
All the aforementioned models use the centered version of $B_j$, which means that the intercept ($\alpha_{V[v]}$ or $\alpha$) represents the expected $\mu_i$ when individual judgments are at average value $\bar{B}$. Centering makes the intercept and main effects in models more interpretable (by shifting the zero point to a more meaningful reference point), but it is not the only way of modeling.

In the model in Equation \ref{eq:with_beta2}, we tried a different approach.

\begin{equation}\label{eq:with_beta2}
    \mu_i = \alpha + \beta_1 B_{j} + \beta_2 V[v] \quad \forall j \in J
\end{equation}

This model does not include an interaction between $B_j$ and $V$, which is fundamentally wrong in the scenario of the \texttt{AITA} subreddit: the final verdict depends on individual judgments expressed before it (the verdict is fundamentally computed based on them). Hence, when measuring the total effect of the final verdict disclosure on individual judgments expressed after, we must take into account the direct effect of variable $B_j$ (individual judgments before) on $V$.

In this case, we did not try a variation of this model where the intercept is stratified by the verdict ($\alpha_{V[v]}$). This is because $V$ (which is a set of dummy variables) is already included in $\beta_2$, and counting it twice would likely lead to multicollinearity (commonly known as the ``dummy variable trap'').


\subsubsection{Non i.i.d. variables}
We need to introduce extra parameters to capture the assumed dependencies.
We have six $\beta$ parameters, one for each possible individual judgment. We define a prior distribution of $B_j$ for each of the six $j$ in $J$. 

\begin{equation}\label{eq:non_iid}
    \mu_i = \alpha{V[v]} + \beta_1 B_{ESH} + \beta_2 B_{NAH} + \beta_3 B_{NTA} + \beta_4 B_{YTA} + \beta_5 B_{unsure} + \beta_6 B_{no\_judg} 
\end{equation}

The results show good convergence (\texttt{r\_hat}=1.0), zero noise in the MCMC estimate, and considerable sample size (between 2,000 and 7,000). However, all the $\alpha$ and the $\beta$ parameters are approximately zero, suggesting an absence of any measurable effect of the final group verdict on individual judgments. This outcome is attributable to degenerate posterior distributions, indicating that the model cannot distinguish between different parameter values due to a lack of underlying variation or structure in the data.

This observation supports the assumption of independence among individual judgments: the value of one variable does not influence the value of any other, implying no systematic relationship or social influence between participants in the dataset. Furthermore, the judgments appear to be identically distributed, meaning that each individual response is generated from the same underlying probability distribution. Consequently, all observations share common statistical properties, suggesting that the process governing individual judgments remains stable across instances.

Although our dataset originates from a time series of posts and responses, analysis reveals that 99\% of users participate only once, and no user alters their judgment during the course of a verdict. This confirms the lack of temporal dependence at the individual level (see Section \ref{subsec:model}).

Thus, the assumption that individual judgments are independent and identically distributed (i.i.d.) is both empirically justified and theoretically sound in this context. Additionally, our data collection can be interpreted as a random sampling of user conversations on the \texttt{AITA} subreddit, further reinforcing the appropriateness of treating individual judgments as i.i.d. observations for the purposes of statistical modeling.


\end{document}